\documentclass[a4paper,11pt]{article}
\usepackage{jheppub} 
\usepackage{lineno}
\usepackage{graphicx} 
\usepackage{amsfonts}
\usepackage{dsfont}
\usepackage{amsmath}
\usepackage{physics}
\usepackage{hyperref}
\usepackage{xcolor}
\usepackage{subcaption}
\usepackage{mdframed}
\usepackage{soul}

\newcommand{\vertextwist}{\alpha}
\newcommand{\edgetwist}{\varepsilon}
\newcommand{\arri}[1]{#1^{\rightarrow}}
\newcommand{\arle}[1]{#1^{\leftarrow}}

\makeatletter
\newcommand\xleftrightarrow[2][]{%
  \ext@arrow 9999{\longleftrightarrowfill@}{#1}{#2}}
\newcommand\longleftrightarrowfill@{%
  \arrowfill@\leftarrow\relbar\rightarrow}
\makeatother

\definecolor{petrolblue}{HTML}{00728D}
\definecolor{darkgray}{HTML}{4D4E4F}
\definecolor{white}{HTML}{FFFFFF}
\definecolor{darkblue}{HTML}{183F56}
\definecolor{lightgray}{HTML}{C8CAD4}
\definecolor{lightblue}{HTML}{9CD7F3}
\definecolor{red}{HTML}{AB3502}
\definecolor{purple}{HTML}{4B70FF}
\definecolor{orange}{HTML}{E69426}

\newcommand{\dint}{\text{d}}

\title{Kinematic Flow for Banana Loops and Unparticles}

\author[a,b]{Tom Westerdijk}
\emailAdd{tom.westerdijk@sns.it}
\author[a,b]{, Chen Yang}
\emailAdd{chen.yang@sns.it}
\affiliation[a]{Scuola Normale Superiore, \\ Piazza dei Cavalieri 7, Pisa, 56126, Italy}
\affiliation[b]{INFN - Sezione di Pisa, \\ Largo Pontecorvo 3, Pisa, 56127, Italy}

\abstract{
We extend kinematic flow to a class of momentum-integrated loop-level cosmological correlators: banana loops of conformally coupled scalars in power-law cosmologies and, in de Sitter, banana loops containing arbitrary mixtures of massless and conformally coupled scalars.
Exploiting their dual description as tree-level exchanges of unparticles, we show that the associated correlators are described by a finite set of master integrals obeying a first-order system of differential equations.
The corresponding basis is constructed from tubings of marked graphs and is distinguished by the appearance of nested tubes and an arborescence ordering of the vertices.
We derive the connection matrices from four combinatorial rules---activation, merger, swap, and copy. The last two are unique to unparticle exchanges: they induce richer mixing among basis functions and introduce new kinematic letters.
Our framework extends systematically to arbitrarily complicated configurations, including necklace diagrams, and establishes unparticle exchange as a distinct class of kinematic flow.
}

\begin{document}

\maketitle

\flushbottom

\section{Introduction}
\label{sec:intro}
Cosmological correlators encode the late-time observables of primordial fluctuations, thereby providing a bridge between fundamental physics in the early universe and measurable quantities such as non-Gaussianities in the cosmic microwave background (CMB). 
Beyond their phenomenological relevance \citep[e.g.,][]{Philcox:2024jpd,Coulton:2024vot,Cabass:2024wob,Sohn:2024xzd,Philcox:2025bvj,Philcox:2025lrr,Philcox:2025wts,Philcox:2026njr}, a growing body of work has revealed that these observables exhibit striking simplicity and organisation, often far beyond what is apparent from their definition in standard perturbation theory. 
This has led to the perspective that cosmological correlators should be viewed not merely as outputs of bulk dynamics, but as objects characterised by intrinsic mathematical data, governed by symmetry, analyticity, and combinatorics.

This paradigm shift has been heavily driven by the cosmological bootstrap program \cite{Arkani-Hamed:2018kmz,Baumann:2019oyu,Baumann:2020dch,Baumann:2022jpr,Pimentel:2022fsc,DuasoPueyo:2023kyh,Aoki:2024uyi}.
The analytic properties of cosmological correlators reveal a web of relations connecting differential equations, singularities, and graphical combinatorics.
Instead of tracking time integrals, one can view these correlators as sources of structured mathematical data whose organisation exposes hidden geometric and algebraic patterns \cite{Fevola:2024nzj}.
Understanding and systematising these structures has therefore become a central theme in recent developments.

A striking example of this emerging structure is \textit{kinematic flow} \cite{Arkani-Hamed:2023kig,Arkani-Hamed:2023bsv,Hang:2024xas,Baumann:2024mvm,Baumann:2025qjx,Glew:2025otn,Glew:2025ugf,Glew:2025ypb,Baumann:2026atn}.
It was observed that in power-law cosmology, a simple set of graphical rules can organise tree-level correlators \cite{Arkani-Hamed:2023kig,Arkani-Hamed:2023bsv} (and later loop integrands \cite{Hang:2024xas,Baumann:2024mvm}) associated with graph tubings, revealing universal patterns underlying their differential equations.
From this perspective, the role of the Feynman diagrams is elevated from a computational tool to a combinatorial object encoding analytic data.
The structure of the differential equations, their singularities, and the space of solutions can all be mapped to graph-theoretic constructions.
In particular, it has been shown that the basis of solutions and their singular loci admit a natural interpretation in terms of tubings of marked graphs, suggesting an underlying geometric organisation closely related to the combinatorics of polytopes and hyperplane arrangements \cite{Arkani-Hamed:2017fdk,Benincasa:2019vqr,Benincasa:2024leu,Arkani-Hamed:2024jbp,Figueiredo:2025daa,Glew:2026von}.
This connection echoes earlier developments in the study of cosmological polytopes and provides further evidence that cosmological correlators admit a geometric reformulation.

From the perspective of bulk perturbation theory, kinematic flow governs the analytic structure of $n$-point correlators organised in terms of in-in Feynman diagrams:
\begin{align}
    \raisebox{-0.45\height}{\includegraphics[width=0.92\linewidth]{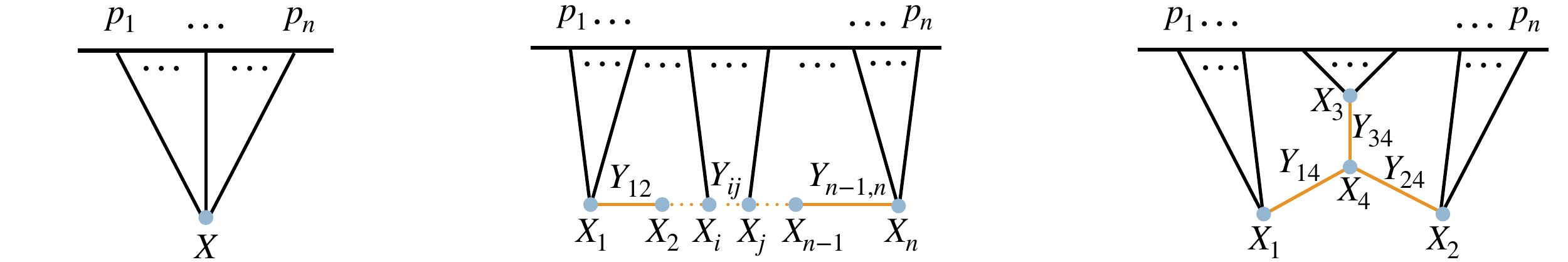}}
    \label{eq:InInCorrelators}
\end{align}

Because time-translation invariance is broken by the cosmological background, every bulk interaction vertex must be integrated over time.
As a stepping stone towards more realistic scenarios, it is often assumed that the external legs connecting to the boundary with a momentum $p_i$ are conformally coupled (CC).
The resulting integrands of the time integrals depend only on the sum of the external energies entering each vertex (denoted as $X_i$), while the internal energies associated with the edges of the graph are denoted as $Y_{ij}$.
All relevant kinematic information is captured by the variables $X_i$ and $Y_{ij}$, and consequently, the analysis only uses graphs where external lines are amputated. 

The existing literature \cite{Baumann:2026atn,Baumann:2025qjx,Arkani-Hamed:2023kig,Arkani-Hamed:2023bsv} has developed kinematic flow for exchanges, denoted by the orange lines in \eqref{eq:InInCorrelators}, of conformally coupled and massive scalars.
These developments naturally raise a question: \textit{how far can the kinematic flow framework be extended?}
Fully integrated cosmological correlators at the loop level introduce complicated new analytic structures and the organisation remains poorly understood.
A promising strategy is to focus on special families of loop diagrams whose structure is sufficiently constrained to admit systematic analysis.
For example, previous works \cite{Cacciatori:2024zrv,Cacciatori:2024zbe} evaluated one- and two-loop (A)dS banana integrals with arbitrary internal masses, contributing to vacuum diagrams and bulk two-point functions in position space. 
Here, instead, we study arbitrary concatenations of banana subdiagrams in power-law cosmologies, at arbitrary loop order and with any number of external legs, as contributions to boundary correlators.
Concretely, we focus on two particularly elegant classes: the banana loop diagrams of an arbitrary number of conformally coupled scalars in general power-law Friedmann-Robertson-Walker (FRW) backgrounds, and ``mixed massless'' banana loops built from massless scalars and conformally coupled scalars in de Sitter.

Remarkably, these banana diagrams admit a dual description in terms of the exchange of \textit{unparticles} \cite{Westerdijk:2025ywh} with integer scaling dimensions.
Unparticles, originally introduced as scale-invariant sectors lacking a traditional discrete particle interpretation, correspond to operators with non-integer scaling dimensions\footnote{Among the phases of IR quantum field theories \cite{Rychkov:2016iqz}, the resulting scale-invariant field is called an \textit{unparticle}. In order to capture the general features of the scale-invariant sector, it is reasonable to regard conformal fields as unparticles. Some phenomenological aspects are discussed in \cite{Strassler:2006im,Cheung:2007zza,Cheung:2007ap,Kikuchi:2007az,Collins:2008ny,Sannino:2008nv,McDonald:2008uh,Sannino:2009za,Fox:2011pm,CMS:2014jvv,Baumgart:2019clc,Pimentel:2025rds,Yang:2025apy,Jiang:2025mlm,Philcox:2026bfa}.}.
Their propagators resemble a continuous superposition of massive modes, producing characteristic behaviours seen in cosmological correlators \cite{Banks:1981nn,Gardi:1998ch,Georgi:2007ek,Georgi:2007si,Stephanov:2007ry,Grinstein:2008qk,Georgi:2009xq}.
By exploiting this duality, one may reinterpret a banana diagram not simply as a loop diagram of fundamental particles, but as the exchange of an effective composite operator whose scaling properties are inherited from the underlying graph \cite{Westerdijk:2025ywh}. 

This unparticle interpretation provides a useful bridge between seemingly complicated loop computations and tree-level operator exchanges.
In this paper, we explore this powerful connection through the lens of kinematic flow.
Our goal is to extend the kinematic flow framework to encompass both banana loops and unparticles, thereby enlarging the class of cosmological correlators whose differential equations can be completely determined by graphical and kinematic principles.

Kinematic flow for unparticles provides a distinct deformation of the conformally coupled base case.
Given an arbitrary Feynman graph, the Feynman rules assign to it an integral $\mathcal I$ computing a contribution to a cosmological correlator.
Each edge is associated with an unparticle propagator, $G_{\text{unparticle}}$, and, in analogy with \cite{He:2024olr,Baumann:2025qjx,Baumann:2026atn}, decomposing this propagator as $G_{\text{unparticle}}\sim G^+ + G^-$ splits $\mathcal I$ into $2^E$ integrals, where $E$ denotes the number of edges.
These $2^E$ functions provide the starting point for a basis of the differential equations, which is then completed by allowing a third local possibility on each edge, giving a total of $3^E$ basis elements.
The additional basis elements play a role analogous to the merger functions in the conformally coupled case.

In the conformally coupled case, the basis admits a graphical description in terms of tubings of marked graphs \cite{Baumann:2025qjx}. A marked graph is obtained by decorating each edge of the underlying Feynman graph with a cross. Tubes are graphical objects that enclose collections of vertices and crosses, while tubings are sets of such tubes. The conformally coupled basis is then described by tube partitions, namely tubings whose constituent tubes cover the full graph.

To extend this construction to unparticle exchange, we equip the graph with an arbitrarily chosen \textit{arborescence} ordering of its vertices. Roughly speaking, this is a partial ordering obtained by picking a root vertex and arranging every other vertex below it. The ordering is purely auxiliary: it simply picks a convenient basis in which the differential equations take their simplest form, and different choices correspond to different but equivalent basis choices.

With this ordering in place, the $3^E$ master integrals are mapped to unparticle tubings, obtained by dressing the conformally coupled tube partitions with $E$ additional \textit{nested} tubes. For every edge not enclosed by a tube, the corresponding cross is itself tubed. For every outer tube that fully contains a subset of edges, one considers all bipartitions of that tube and, from each partition, selects the lower tube according to the arborescence ordering. These selected tubes are then added as nested tubes. 
\begin{figure}[h!]
  \centering
  \begin{subfigure}[c]{0.19\textwidth}
    \centering
    \includegraphics[height=1.1cm]{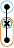}
  \end{subfigure}
  \hfill
  \begin{subfigure}[c]{0.19\textwidth}
    \centering
    \includegraphics[height=1.1cm]{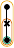}
  \end{subfigure}
  \begin{subfigure}[c]{0.19\textwidth}
    \centering
    \includegraphics[height=1.8cm]{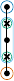}
  \end{subfigure}
  \begin{subfigure}[c]{0.19\textwidth}
    \centering
    \includegraphics[height=1.8cm]{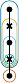}
  \end{subfigure}
  \begin{subfigure}[c]{0.19\textwidth}
    \centering
    \includegraphics[height=1.5cm]{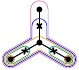}
  \end{subfigure}
  \caption{Five examples of unparticle tubings.}
  \label{fig:main}
\end{figure}

The power of this graphical language lies in the fact that tubings automatically encode information about the differential equations they obey.
Each tube in a tubing represents a \textit{letter}, namely a linear function that is the argument of a $\dint\log$ form appearing in a system of differential equations.
By a slight abuse of terminology, we also call the $\dint\log$ forms of linear functions letters. 
For a given basis function, the set of letters is exactly the set of tubes contained in its associated tubing.
For each of these letters, the kinematic flow rules determine the corresponding coefficient as a sum of basis functions weighted by \textit{twist} parameters, which encode the spacetime and the scaling dimensions.

The differential equations for unparticle exchange are determined by four kinematic flow rules: \textit{activation}, \textit{merger}, \textit{swap}, and \textit{copy}.
The \textit{activation rule} contributes the differentiated function itself, multiplied by the sum of all letters associated with the tubes in its tubing.
The coefficients are fixed by the twist parameters. Nested tubes come with edge parameters, $\edgetwist_{ij}$.
Tubes that are not nested carry sums of vertex twists, $\alpha_i$, and also receive $\edgetwist_{ij}$ contributions from edges that enter from below.

The other three rules generate new basis functions.
The \textit{merger rule} forms a new tubing by merging two adjacent tubes in the differentiated one and then dressing the result with nested tubes.
Its associated letters are those of the two tubes that are merged.
The \textit{swap rule} applies to adjacent tubes as well: it swaps a cross on the marked graph and in this way produces another tubing.
The associated letters are the tubed cross and, among the two adjacent tubes, the uppermost one selected by the arborescence ordering.
Finally, when nested tubes are present, the \textit{copy rule} produces a new tubing by cutting the differentiated tubing along a nested tube.
In this case, the differential equation does not involve the corresponding function itself.
Instead, one takes the differential of the cut function on the nested letter under consideration, or equivalently copies all functions multiplying that nested letter in the differential equation of the cut tubing.
Their sum is then multiplied by the difference between the nested letter and the letter of the surrounding outer shell.

In this article, we start by concisely reviewing the studied class of cosmological correlators and how to build the corresponding integrals in bulk perturbation theory.
We emphasise the key role of the unparticle propagator, which describes not only scale-invariant sectors but also important classes of loop diagrams.
After giving a concrete formula for the studied family of integrals, we switch gears and precisely define the basis of functions in terms of tubings.
In section~\ref{sec:letters}, we explain the conventional setup in detail, and then use this to construct unparticle tubings introducing the concepts of nested tubes and arborescence ordering.
Building upon the definition of the bases and the graphical language, we describe the four kinematic flow rules in section~\ref{sec:rules} and illustrate their application with a handful of examples in section~\ref{sec:examples}.

\paragraph{Setup.}
We are interested in equal-time correlators of conformally coupled scalars that exchange an arbitrary number of unparticles in an FRW universe.
The flat-sliced FRW metric for a power-law cosmology is 
\begin{align}
    \dint s^2 = \frac{-\dint\eta^2+\dint\vec{x}^2}{(\eta/\eta_0)^{2(1+\delta)}} \equiv a(\eta)^2 \left(-\dint\eta^2+\dint\vec{x}^2\right). 
    \label{def:frw_metric}
\end{align}
Throughout this work, we work with $\eta_0\equiv-1$.
We use dimensional regularisation (dim-reg) to regularise the UV divergences of the loops: $D\equiv d+1=4+2\epsilon$.
Notice that here $\epsilon$ is a complex parameter.

Conformally coupled scalars, denoted by $\varphi$, have a time-dependent mass that is generated by a coupling to the Ricci scalar, with polynomial interactions: 
\begin{align}
    S_\varphi = -\int\dint^Dx\ \sqrt{-g} \left(\frac{1}{2}\partial_\mu\varphi\partial^\mu\varphi + \xi R \varphi^2 + \sum_n \lambda_n \frac{\varphi^n}{n!}\right). 
\end{align}
Here, the coupling constant is $\xi\equiv\frac{D-2}{4(D-1)}$.
The mode function of conformally coupled scalars in the Bunch-Davies vacuum is 
\begin{align}
    f_p(\eta) = (-\eta)^{(1+\delta)(1+\epsilon)} \frac{e^{-ip\eta}}{\sqrt{2p}}, 
\end{align}
where $p\equiv|\vec{p}|$ is the magnitude of the three-dimensional momentum.

\paragraph{Feynman rules.}
We compute equal-time correlators in the ``in-in'' formalism.
The perturbative calculation of these correlators can be organised in terms of Feynman diagrams.
Vertices in these diagrams can be either on the $+$ or $-$ in-in branch.

\begin{figure}[b]
    \centering
    \includegraphics[width=0.7\linewidth]{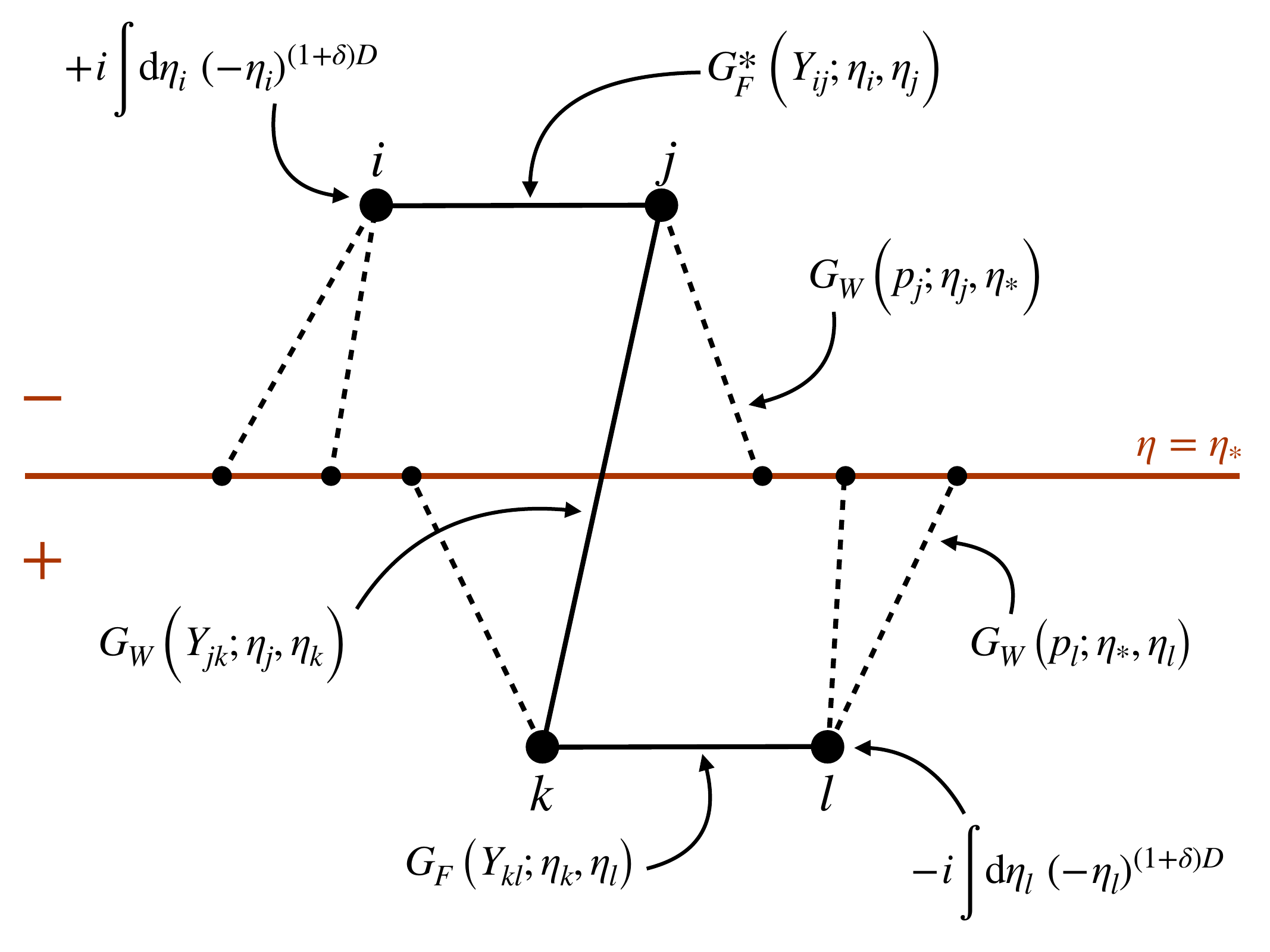}
    \caption{The Feynman rules.}
    \label{fig:feynman_rules}
\end{figure}

Suppose that we are given a Feynman graph, $G$, with a set of vertices $V$ and internal edges $E$.
Consider all partitions of $V$ into vertices on the $+$ branch collected in $V^+$ and on the $-$ branch in $V^-$.
The Feynman rules for in-in correlators \cite{Weinberg:2005vy,Giddings:2010ui} assign an integral formula to the graph $G$ with a partitioned vertex set.
For simplicity we set all coupling constants to one.
In figure~\ref{fig:feynman_rules}, we present the rules by applying them to a concrete example.
Vertices correspond to time integrals equipped with the right measure and a crucial sign difference between the two branches.
An edge corresponds to a time-ordered propagator, $G_F$ ($G_F^*$), if it connects two vertices on the $+$ ($-$) branch.
External edges and edges that connect two vertices on different branches correspond to Wightman functions whose leftmost time argument corresponds to the upper vertex.
The total correlator is obtained by summing over all possible partitions of $V$ into $V^+$ and $V^-$.

Throughout this paper the external legs will always be composed of conformally coupled scalars and thus 
\begin{align}
    G_W(p;\eta_*,\eta) \equiv f_p(\eta_*)f_p^*(\eta) \overset{\rm CC}{\propto} (-\eta)^{(1+\delta)(1+\epsilon)} \, e^{ip\eta}.
\end{align}
For multiple external legs attached to a vertex $v_k$, the resulting product of Wightman functions is functionally equivalent to a single Wightman function: 
\begin{align}
    (-\eta_k)^{N_k(1+\delta)(1+\epsilon)}\, e^{i X_k\eta_k},
\end{align}
where $N_k$ is the number of external legs and $X_k=\sum_{i=1}^{N_k}p_i$.
By virtue of this property, we can truncate the external legs and work with graphs solely composed of the vertices and internal edges.

Before diving into unparticles, loops and the associated propagators, let us mention that we will only be concerned with deriving differential equations for the correlator from a single sector: the all $+$ sector where every vertex is on the $+$ branch.
In general, different sectors in the in-in formalism are related by complex conjugation (sometimes called a ``reflection symmetry'') which exchanges vertices between the $+$ and $-$ branches, effectively reflecting about the $\eta=\eta_*$ surface (see figure~\ref{fig:feynman_rules}).

Furthermore, the rest of the independent sectors can be obtained by replacing sets of $G_F(Y_{ij};\eta_i,\eta_j)$ propagators with their $G_W(Y_{ij};\eta_i,\eta_j)$ counterparts and flipping the sign of $X_i$.
For the special case of unparticles, the Wightman function satisfies exactly the same differential equation as the time-ordered propagator.
Consequently, by starting with $G_F(Y_{ij};\eta_i,\eta_j)$ in the integrand, and then moving a vertex $v_i$ from the $+$ branch to the $-$ branch only results in $X_i \rightarrow -X_i$.
For the two-site case, this feature was already observed in \cite{Westerdijk:2025ywh}, where it made the derivation of the differential equations for the $+-$ sector redundant.
For more complicated graphs, this property reduces the calculation of the plethora of sectors to just a single sector which we take to be the all $+$ sector.

\vspace{1em}
\noindent\textbf{Unparticles.}
It is possible that in the UV completion for the EFT of inflation, there are fields that become non-trivially scale-invariant at the Hubble scale during inflation.
As a simple model, we treat these scale-invariant sectors as conformal field theories (CFTs).
For a conformal field, the scalar two-point function in Euclidean flat space is completely fixed by conformal symmetry,
\begin{align}
    \langle \mathcal{O}_\Delta (\tau_1,\vec{x}_1)\mathcal{O}_\Delta (\tau_2,\vec{x}_2) \rangle_{\text{flat}} = \frac{1}{\Big((\tau_1 - \tau_2)^2 + (\vec{x}_1-\vec{x}_2)^2\Big)^\Delta}. 
    \label{eq:CFTflat}
\end{align}
Since the FRW metric \eqref{def:frw_metric} is conformally flat, the corresponding two-point function in FRW spacetime can be obtained by performing a Wick rotation $\tau\to i\eta$ and rescaling the operators, $\mathcal{O}_\Delta(x) \rightarrow \tilde{\mathcal{O}}_\Delta(a(\eta) x)=a(\eta)^{-\Delta} \mathcal{O}_\Delta$: 
\begin{align}
    \langle \tilde{\mathcal{O}}_\Delta (\eta_1,\vec{x}) \tilde{\mathcal{O}}_\Delta (\eta_2,\vec{y}) \rangle_{\text{FRW}} &= \frac{(\eta_1\eta_2)^{\Delta(1+\delta)}}{\Big(-(\eta_1 - \eta_2)^2 + (\vec{x}-\vec{y})^2\Big)^\Delta}.
    \label{eqn:2pt_frw}
\end{align}
We need to respect the unitarity bound for scalar conformal fields in $D$ dimensions, $\Delta\geq(D-2)/2$, assuming the unparticle here is a scalar primary operator.

\paragraph{Banana loops as unparticles.}
In a power-law cosmology, the $n$-banana loops of conformally coupled scalars are equivalent to unparticle two-point functions with integer scaling dimension $\Delta=n(1+\epsilon)$ \cite{Westerdijk:2025ywh}.
This follows straightforwardly from the composition of $n$ position-space two-point functions of a conformally coupled scalar. In $d=3+2\epsilon$ dimensions such a two-point function reads 
\begin{align}
    G(\eta_1,\vec x;\eta_2,\vec y)=\frac{\, \Gamma(1+\epsilon)}{4\pi^{2+\epsilon}} \left( \frac{(\eta_1\eta_2)^{1+\delta}}{-(\eta_1-\eta_2)^2+(\vec x-\vec y)^2} \right)^{1+\epsilon}.
\end{align}
Here we generalise this correspondence to the ``mixed massless'' banana loops in de Sitter, where $\delta=0$.
We can write a general formula for the position space propagator of fields with $\sqrt{(d/2)^2-m^2}\equiv n+1/2$ in terms of a definite sum\footnote{We thank Harry Goodhew for sharing this observation with us.}:
\begin{align}
    G_{n}(\sigma)=-\frac{\csc(\pi\epsilon)}{4}(4\pi\sigma)^{-1-\epsilon}\sum_{a=0}^n \frac{(n+a)!}{a!(n-a)!}\frac{(-\sigma)^a}{\Gamma(a-\epsilon)},
    \label{eq:DiscreteSeriesProp}
\end{align}
where $\sigma$ is the invariant de Sitter distance
\begin{align}
    \sigma \equiv \frac{-(\eta_1-\eta_2)^2+(\vec x-\vec y)^2}{4\eta_1\eta_2}.
\end{align}
This class of scalar particles belongs to the discrete series of irreducible representations of de Sitter and includes conformally coupled ($n=0$) and massless scalars ($n=1$).

For conformally coupled scalars, the position space propagator is simple:
\begin{align}
    G_{n=0}(\sigma)=\frac{\Gamma(1+\epsilon)}{4\pi}(4\pi\sigma)^{-1-\epsilon}.
\end{align}
For massless scalars, the position space propagator has two terms:
\begin{align}
    G_{n=1}(\sigma)=\frac{\Gamma(1+\epsilon)}{4\pi}(4\pi)^{-1-\epsilon}\left(\sigma^{-1-\epsilon}+\frac{2}{\epsilon}\sigma^{-\epsilon}\right).
\end{align}

A loop built from a product of propagators of the form \eqref{eq:DiscreteSeriesProp} remains a sum of powers of $\sigma$. Every term in this sum can be thought of as an unparticle exchange with a certain $\epsilon$-corrected integer scaling dimension.
For a mixed massless loop with $p$ conformally coupled scalars and $q$ massless scalars, the composition of two-point functions will result in
\begin{align}
    G_{\rm eff}
    &= G_{n=0}^p G_{n=1}^q \\
    &= \left(
        \frac{\Gamma(1+\epsilon)}{4\pi}
        (4\pi)^{-1-\epsilon}
       \right)^{p+q}
       \sigma^{-(p+q)(1+\epsilon)}
       \sum_{k=0}^{q}
       \begin{pmatrix} q \\ k \end{pmatrix}
       \left(\frac{2\sigma}{\epsilon}\right)^k\,.
    \label{eq:Geff}
\end{align}
The sum in \eqref{eq:Geff} arises from a binomial expansion and represents a linear combination of unparticle propagators with different scaling dimensions.
Specifically, the scaling dimension will take the form $\Delta=(p+q)(1+\epsilon)-k$, in which $k$ is an integer from $0$ to $q$.
It is worth mentioning that, when the banana loop is fully composed of massless scalars, the unparticle expansion contains one term with scaling dimension $\Delta=q\epsilon$, which breaks the scalar unitarity bound while staying positive.
This violation of the unitarity bound signals the IR divergences in massless banana loops, as also observed in \citep[e.g.,][]{Senatore:2009cf,Andrade:2011dg,Senatore:2012nq,Cespedes:2023aal}.
Despite these IR divergences, the formalism still applies and captures important features of massless banana loops.

In other words, any diagram built from these ``mixed massless'' two-site loops decomposes into a sum of unparticle exchanges if $\delta=0$, as indicated below, where the wavy and solid lines correspond to massless and conformally coupled scalars, respectively:
\begin{align}
    \raisebox{-0.5\height}{\includegraphics[width=0.3\linewidth]{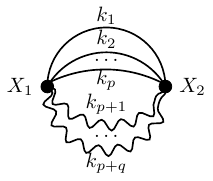}} \equiv \sum_{k=0}^{q} c_k \ \raisebox{-0.4\height}{\includegraphics[width=0.47\linewidth]{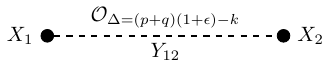}}
    \label{eqn:bananas_are_unparticles}
\end{align}
The coefficients $c_k$, which can be derived from \eqref{eq:Geff}, depend on $p, \ q$, and $\epsilon$, . 

\paragraph{In-in unparticle propagators.}
In inflationary cosmology, since we are computing equal-time correlators on the late-time boundary using the ``in-in" formalism, we go to spatial momentum space.
Fourier transforming the left-hand side in $d$ dimensions gives
\begin{align}
    \int \dint^{d} x\ e^{-i\vec{Y}\cdot\vec{x}}\frac{1}{(-\eta^2+\vec x^2)^\Delta} &= \frac{(2\pi)^{\frac{d+1}{2}}}{4^{\Delta-D/4}\Gamma(\Delta)\sqrt{\pi}} \cdot (2Y)^{\Delta - \frac{d}{2}} \cdot \frac{1}{\abs{\eta}^{\Delta - \frac{d}{2}}} \cdot K_{\Delta - \frac{d}{2}}\Big(iY \abs{\eta}\Big).
    \label{eqn:2pt_besselK_int_repre}
\end{align}
In $d+1$-dimensional FRW space, after performing the same transformation as in \eqref{eqn:2pt_frw} and normalising by a factor of $(2\pi)^{\frac{d+1}{2}}$, we can see that the time-ordered propagator for unparticles is
\begin{align}
    G_{F}(Y;\eta_1,\eta_2) = 
    \frac{(\eta_1\eta_2)^{\Delta(1+\delta)}}{2^{2\Delta+\frac{1-d}{2}}\sqrt{\pi}\,\Gamma(\Delta)}\, U_{\Delta-\frac{d}{2}}(Y;|\eta_1-\eta_2|),
    \label{def:deSitter_G_F} 
\end{align}
and the Wightman function,
\begin{align}
    G_{W}(Y;\eta_1,\eta_2) = 
    \frac{(\eta_1\eta_2)^{\Delta(1+\delta)}}{2^{2\Delta+\frac{1-d}{2}}\sqrt{\pi}\,\Gamma(\Delta)}U_{\Delta-\frac{d}{2}}(Y;\eta_1-\eta_2),
\end{align}
where we have defined
\begin{align}
    U_{\Delta-\frac{d}{2}}(Y;\eta)
    :=
    \left( \frac{2 Y}{i\eta}\right)^{\Delta-\frac{d}{2}}K_{\Delta-\frac{d}{2}}\big(i Y\eta\big).
\end{align}

If we take $\Delta=1+\epsilon$, \eqref{def:deSitter_G_F} will reduce to the known propagator of conformally coupled scalars.

\paragraph{Master integrals.}
An initial set of master integrals whose sum gives the connected correlator $C_G$ is constructed using a new set of Feynman diagrams and rules.
This set does not close under differentiation; we will complete the basis in the next section.
In a similar spirit to \cite{Baumann:2025qjx,Baumann:2026atn}, we decompose every edge into two new types of edges, $+$ and $-$.
This edge-labelling should \textit{not} be confused with the in-in branches.
In the new Feynman rules, they are associated with
\begin{equation}
    G^\pm(Y;\eta_1,\eta_2) = \frac{(\eta_1\eta_2)^{\Delta(1+\delta)}}{2^{2\Delta+\frac{1-d}{2}}\sqrt{\pi}\,\Gamma(\Delta)}\, U^\pm_{\Delta-\frac{d}{2}}(Y;|\eta_1-\eta_2|)
    \label{eq:GpmDef}
\end{equation}
where
\begin{align}
    U^\pm_{\Delta-\frac{d}{2}}(Y;\eta)=\frac{1}{2}(Y\pm i\partial_{\eta})U_{\Delta-\frac{d}{2}}(Y;\eta).
\end{align}
The $\pm$ propagators satisfy the essential property that
\begin{align}
    G_F(Y;\eta_1,\eta_2)= \frac{1}{Y} \times \left(G^+(Y;\eta_1,\eta_2)+G^-(Y;\eta_1,\eta_2)\right).
    \label{eq:GFinPM}
\end{align}

Adopting the same notation as before, $V$ is the set of vertices and $E$ the set of edges of a graph $G$.
We now further partition the set $E$ into a set of $+$ edges, $E^+$, and a set of $-$ edges, $E^-$.
Neglecting the universal numerical and $d$-dependent factors, the master integral for a particular partition reads
\vspace{-1.5em}
\begin{center}
    \fbox{\begin{minipage}{.99\textwidth}
        \begin{align}
            \mathcal I(V,E^+,E^-)=\prod_{v_k \in V}\int \dint \eta_k (-\eta_k)^{-1-\alpha_k} \, e^{i X_k\eta_k}  \prod_{e_{mn}\in E^+} G_{mn}^+ \prod_{e_{mn}\in E^-}G^-_{mn},
            \label{eq:StudiedIntegrals}
        \end{align}
    \end{minipage}}
\end{center}
where every propagator has a different scaling weight
\begin{align}
    G^\pm_{mn}\equiv G_{\Delta_{mn}}^\pm(Y_{mn};\eta_m,\eta_n).
\end{align}
The connected correlator can be reconstructed from the set of master integrals by summing over all edge partitions:
\begin{equation}
    C_G(V,E)=\prod_{e_{mn}\in E}Y_{mn} \sum_{E^+,E^-} \mathcal I(V,E^+,E^-) \quad \text{with} \ E^+\cup E^- = E \ \text{and} \ E^+\cap E^- = \emptyset.
\end{equation}
In other words, this is what one obtains by expanding the product of all \eqref{eq:GFinPM} once for each edge.
Let $N_k$ be the number of external legs connected to the vertex $v_k$.
The ``twists'' of the time integrals can be written in closed form:
\begin{align}
    \alpha_k = -1-(1+\delta)\left[ \sum_l \Delta_{kl}+N_k\frac{d-1}{2}-(d+1) \right],
    \label{eq:VertexTwist}
\end{align}
where we are summing over all edges, $e_{kl}$, that are connected to $v_k$.
The parameter $\alpha_k$ appears linearly in the differential equations.
The twist associated with the scaling weight $\Delta_{kl}$ of each internal edge also appears linearly, and is given by
\begin{equation}
    \edgetwist_{kl}=\Delta_{kl}-\frac{d-1}{2}.
    \label{eq:EdgeTwist}
\end{equation} 

\section{Kinematic flow for banana loops and unparticles}
\label{sec:flow}
Kinematic flow provides a structured method to derive the connection matrices, $A$, for a system of first-order differential equations,
\begin{align}
    \dint \vec{\mathcal I} = A\cdot \vec{\mathcal I},
    \label{eq:Basic_DE}
\end{align}
where the vector $\vec{\mathcal I}$ collects a basis of master integrals.
The central idea is to encode the differential structure directly at the level of graph-based objects, bypassing the need for explicit integral manipulations at intermediate steps.

In this section, we first introduce the graphical language by explaining the concepts of marked graphs and tubes in detail. Using the conformally coupled case as a starting point, we show how to construct the tubings associated with the unparticle basis functions bottom-up via arborescence ordering and nested tubes. With the basis in terms of tubings in hand, we then connect this graphical picture in a straightforward manner to the integrals in \eqref{eq:StudiedIntegrals} that sum up to the correlator. We also describe three essential graph operations: merger, cut and swap. This sets the stage for the kinematic flow rules which we formulate precisely before illustrating the rules with a handful of examples.

\subsection{Letters and master integrals}
\label{sec:letters}
For a given Feynman graph, $G$, we describe here how to construct the master integrals, i.e. $\vec{\mathcal I}$, for the fully time-ordered sector.
Everything will be formulated solely in terms of decorated graphs.
We have separated the construction into two parts: first, we detail the conformally coupled case, after which we explain how to add the nested structure that arises for unparticles.

\paragraph{Graphs, tubes, and letters.}
Consider an arbitrary tree graph and call it $G$.
It contains $V$ vertices denoted by $v_i$ with $i=1,\ldots,V$ and $(V-1)$ edges denoted by $e_{ij}$ each connecting two vertices $v_i$ and $v_j$.
Every edge is decorated with a cross, $\times_{ij}$.
\begin{align}
    \raisebox{-0.5\height}{\includegraphics[width=0.2\linewidth]{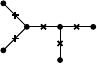}}
\end{align}

To classify the letters and master integrals, we will need the concept of a tube.
A tube, $T$, is a subset of vertices and crosses in $G$ that makes up a connected subgraph of $G$.
Consider a vertex-induced connected subgraph, $H \subseteq G$, and its boundary, $\partial H$, consisting of all edges that connect $H$ with $G \setminus H$.
We define a tube as the triplet of the subgraph, its boundary, and a subset of its boundary $T=(H,\partial H,B \subseteq \partial H)$. In other words, a tube is a collection of vertices that make up a connected subgraph of $G$ combined with a subset of edges that connect the subgraph with the rest of $G$.

Graphically, we can represent this as a tube enclosing a subset of vertices and crosses so that everything inside the tube is connected.
The included vertices define the subgraph; the contained crosses indicate which edges are in $B$.
Importantly, we distinguish between tubes not only by the vertices of $G$ they contain but also by the crosses they enclose.
They have an easy graphical representation:
\begin{align}
    \raisebox{-0.5\height}{\includegraphics[width=0.2\linewidth]{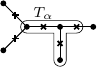}} \hspace{0.1\linewidth} 
    \raisebox{-0.5\height}{\includegraphics[width=0.2\linewidth]{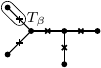}}
\end{align}

The triplets for $T_\alpha$ and $T_\beta$ can be read off from the above.
Pictorially, for example for $T_\alpha$:
\begin{align}
    H_\alpha = \raisebox{-0.55\height}{\includegraphics[width=0.2\linewidth]{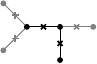}},\ \partial H_\alpha = \raisebox{-0.55\height}{\includegraphics[width=0.2\linewidth]{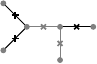}},\ B_\alpha = \raisebox{-0.55\height}{\includegraphics[width=0.2\linewidth]{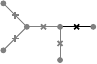}}
\end{align}

To write down a formula for every letter that can appear in the differential equations, we relate tubes to linear functions of the kinematic variables.
Let us first assign to each vertex labelled by $i$ the vertex-energy $X_i$ and to each edge connected to vertices $i$ and $j$, the edge-energy $Y_{ij}$.
Given a tube $T$, the associated letter is
\begin{align}
    L(T) &= \sum_{v_i\in H} X_i + \sum_{e_{ij}\in \partial H\setminus B}Y_{ij} - \sum_{e_{ij}\in B} Y_{ij}.
    \label{eq:LetterMap}
\end{align}
For example,
\begin{align}
    \raisebox{-0.5\height}{\includegraphics[width=0.2\linewidth]{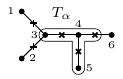}} &\longrightarrow L(T_\alpha) = X_3+X_4+X_5+Y_{13}+Y_{23}-Y_{46}, \\
    \raisebox{-0.5\height}{\includegraphics[width=0.2\linewidth]{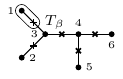}} &\longrightarrow L(T_\beta) = X_1-Y_{13}.
\end{align}
As a special case where we only include a single cross, we specify
\begin{align}
    L(\times_{ij})=Y_{ij},
\end{align}
or equivalently $T(\times_{ij})=(\emptyset,e_{ij},\emptyset)$.
Graphically, this special case is represented by encircling the cross under consideration.
Here is an example:
\begin{align}
    \raisebox{-0.5\height}{\includegraphics[width=0.2\linewidth]{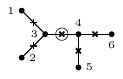}} &\longrightarrow L(\times_{34}) = Y_{34}.
\end{align}
We can distinguish between two types of tubes:
\begin{itemize}
    \item \textit{Conformally coupled.}
    In this case, one needs to additionally demand that the tube includes at least one vertex of $G$: $H \neq \emptyset$.
    This simply means that the tube cannot include only a single cross.
    We call such a tube a \textit{CC tube}.
    The set of letters in the conformally coupled differential equation corresponds exactly to the set of CC tubes one can draw on the marked graph $G$.
    \item \textit{Unparticles.}
    In addition to the CC tubes, one can also tube a single cross, i.e., take $\partial H =e_{ij}$ while $H=B=\emptyset$.
    Therefore, the set of letters for unparticles is supplemented with $Y_{ij}$ for all $e_{ij} \in G$.
\end{itemize}

\paragraph{Master integrals, conformally coupled.}
The master integrals are uniquely determined by the letters that appear in their differential equations.
In other words, we can uniquely specify a master integral by drawing a set of tubes, that is, a tubing on the marked graph. 

The conformally coupled tubings of the fully time-ordered sector (equivalently, where all vertices lie on the $+$ branch of the in-in formalism) are obtained by considering all CC tube partitions of the graph $G=(V(G),E(G))$: a set of CC tubes, $T_i=(H_i,\partial H_i,B_i)$ such that their union gives the full graph, $\bigcup_i V(H_i)=V(G)$ and $\bigcup_i(E(H_i)\cup B_i)=E(G)$, and there are no pairs with overlap: $H_i \cap H_j=\emptyset$ and $B_i \cap B_j = \emptyset$ for all $i$ and $j$.
Equivalently, we split the set of vertices and crosses up into $\leq V$ subsets of vertices and crosses that comprise CC tubes: all contained vertices are connected by edges and there is at least one vertex in the subgraph.
\begin{align}
    \raisebox{-0.5\height}{\includegraphics[width=0.2\linewidth]{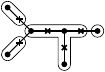}} \hspace{0.1\linewidth} 
    \raisebox{-0.5\height}{\includegraphics[width=0.2\linewidth]{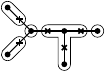}}
\end{align}

\paragraph{Unparticle master integrals.}
The size of the conformally coupled basis in the connected sector (all $+$ in-in sector) is fixed by the three possible ways in which each cross can be tubed: it may belong only to the tube on the left, only to the tube on the right, or to both. Since this choice is made independently for each of the $E$ edges, the total number of master integrals is $3^E$. Unparticle tubings are obtained by supplementing the conformally coupled tubings with $E$ additional tubes. As a result, the number of master integrals in the connected sector is exactly the same for unparticle exchange as for conformally coupled exchange.

The additional \textit{dressing} with $E$ tubes follows two basic rules: for unmerged edges, tube the corresponding cross, and inside a CC tube $T$ with $n$ merged edges, nest $2n$ tubes according to the \textit{arborescence rule}. 
To apply the arborescence rule, we first organise the graph into an \textit{arborescence} by choosing a distinguished \textit{root} vertex. In such a graph, each vertex may have an arbitrary number of incoming edges, while every other vertex has exactly one outgoing edge and the \textit{root} has none\footnote{Strictly speaking, this description corresponds to an anti-arborescence.}.
Crucially, every connected subgraph of an arborescence is an arborescence itself.

When drawing graphs, we follow the convention that the ordering is directed upwards.
A Feynman graph can be algorithmically transformed into an arborescence: we first promote an arbitrary vertex in the graph to the root and then order the rest of the vertices below it.
Within each subgraph that is connected to the root, we repeat the ordering procedure, treating the vertex connected to the root as the local root in the subgraph.
In \eqref{eqn:tree_ordering_illustration}, we organise an example of a graph with its CC tubing into an arborescence.
\begin{align}
    \raisebox{-0.5\height}{\includegraphics[width=0.2\linewidth]{tubings/illustration/illus_5.1.pdf}} &\longrightarrow \raisebox{-0.45\height}{\includegraphics[width=0.2\linewidth]{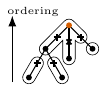}}
    \label{eqn:tree_ordering_illustration}
\end{align}

Importantly, the arborescence ordering is assigned globally to a graph $G$ and is inherited by every tubing of $G$. We can now state the dressing rules in full: 
\vspace{.4em}

{
\centering
\fbox{
\begin{minipage}{.9\textwidth}
\vspace{.5em}
        \begin{enumerate}
            \item Tube every cross if its corresponding edge is not enclosed by a tube.
            \item Nest $2n$ tubes inside every outer tube enclosing $n$ edges by considering the $2n$ bipartitions of the outer tube into two other CC tubes and selecting from each partition the lower tube.
            \vspace{.5em}
        \end{enumerate}
\end{minipage}\hspace{.04\textwidth}}
\par
}

\vspace{.4em}
To be more precise, every pair of tubes considered in the partitions of the second rule share an edge which they both cross. This edge is connected to two vertices, $v_1$ and $v_2$, and they are ordered according to the arborescence prescription, without loss of generality: $v_1 > v_2$. The lowest-lying tube is the tube that contains the lowest lying vertex, $v_2$.

If we draw the graph with its vertices ordered vertically, it is straightforward to select the lowest-lying tube.
Below we give an illustration of two bipartitions into CC tubes where the arrows point to the tube that is selected by the arborescence ordering:
\begin{align*}
    \raisebox{-0.5\height}{\includegraphics[width=0.15\linewidth]{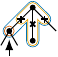}} \hspace{0.1\linewidth} 
    \raisebox{-0.6\height}{\includegraphics[width=0.15\linewidth]{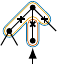}}
\end{align*}
One has to repeat this procedure for every possible bipartition of the outer shell.

In general, this will create tubings with nested tubes, but not intersecting tubes of the kind that appeared in \cite{Baumann:2026atn}.
As another consequence of the construction, if the outer CC tube contains $n$ edges, the number of nested tubes is $2n$.
Edges enclosed by a tube were interpreted in \cite{Baumann:2025qjx} as collapsed internal lines. Here, the additional dressing in the form of nested tubes does not allow for a clean physical interpretation. This is confirmed by the bulk derivation of the rules in appendix~\ref{app:proof_rules} where the mergers arise as auxiliary functions to construct a Pfaffian system.

The arborescence is an auxiliary device that only serves to define a convenient basis of master integrals resulting in the most compact form of the differential equations. Different choices of arborescence generally lead to different tubings and hence to different connection matrices, but these are simply different representations of the same system. In particular, the physical correlator associated with the original graph is independent of this choice.

\paragraph{Examples.}
These rules allow us to completely determine all master integrals that are required to describe the differential equations corresponding to a certain graph $G$.
Let us now explicitly list all tubings for a few simple graphs.
We associate to each vertex and edge a twist parameter:
\begin{align*}
    \raisebox{-0.6\height}{\includegraphics[height=1.5cm]{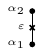}}
    \hspace{0.1\textwidth}
    \raisebox{-0.6\height}{\includegraphics[height=6em]{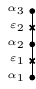}}
    \hspace{0.1\textwidth}
    \raisebox{-0.6\height}{\includegraphics[height=4.8em]{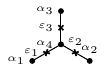}}
\end{align*}
\begin{itemize}
    \item \textit{Two-site:} we choose the root of the arborescence to be vertex $v_2$ (labelled by the parameter $\alpha_2$ above).
    Although this example is fairly simple, we stick to the convention of drawing the root at the top.
    The three functions represented by their respective graph tubings are
    \begin{align}
            \raisebox{-0.5\height}{\includegraphics[height=1.5cm]{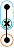}}
        \hspace{0.1\textwidth}
            \raisebox{-0.5\height}{\includegraphics[height=1.5cm]{tubings/two/two_2.pdf}}
        \hspace{0.1\textwidth}
            \raisebox{-0.5\height}{\includegraphics[height=1.5cm]{tubings/two/two_3.pdf}}
    \end{align}
    \item \textit{Three-site:} again, we choose vertex $v_1$ to be the root of the arborescence which forces a complete ordering: $v_3<v_2<v_1$.
    As in the conformally coupled case, there are nine master integrals in total:
    \begin{align}
        \begin{aligned}
            &\raisebox{-0.5\height}{\includegraphics[height=4.8em]{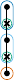}}
            \hspace{0.075\textwidth}
            \raisebox{-0.5\height}{\includegraphics[height=4.8em]{tubings/three/three_2.pdf}}
            \hspace{0.075\textwidth}
            \raisebox{-0.5\height}{\includegraphics[height=4.8em]{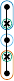}}
            \hspace{0.075\textwidth}
            \raisebox{-0.5\height}{\includegraphics[height=4.8em]{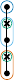}}
            \hspace{0.075\textwidth}
            \raisebox{-0.5\height}{\includegraphics[height=4.8em]{tubings/three/three_5.pdf}} 
            \hspace{0.075\textwidth}
            \raisebox{-0.5\height}{\includegraphics[height=4.8em]{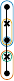}}
            \hspace{0.075\textwidth}
            \raisebox{-0.5\height}{\includegraphics[height=4.8em]{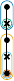}}
            \hspace{0.075\textwidth}
            \raisebox{-0.5\height}{\includegraphics[height=4.8em]{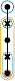}}
            \hspace{0.075\textwidth}
            \raisebox{-0.5\height}{\includegraphics[height=4.8em]{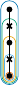}}
        \end{aligned}
    \end{align}
    \item \textit{4-site star examples:} in total, the basis consists of $27$ functions of which we present four examples below:
    \begin{align}
        \raisebox{-0.5\height}{\includegraphics[height=4.8em]{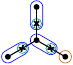}}
        \hspace{0.09\textwidth}
        \raisebox{-0.5\height}{\includegraphics[height=4.8em]{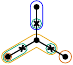}}
        \hspace{0.09\textwidth}
        \raisebox{-0.5\height}{\includegraphics[height=4.8em]{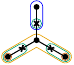}}
        \hspace{0.09\textwidth}
        \raisebox{-0.5\height}{\includegraphics[height=4.8em]{tubings/star/star_27.pdf}}
    \end{align}
\end{itemize}

\paragraph{Relation to time integrals.}
Ultimately, we want to study the integrals introduced in \eqref{eq:StudiedIntegrals}. The detailed relation between the tubings and the time integrals is spelled out in appendix~\ref{app:proof_rules}, which also {details a derivation} of the kinematic flow rules. Here, we are merely concerned with a practical map between the two. The time-integrals were naturally formulated in terms of graphs consisting of vertices and two types of edges: $e^+_{ij}$ and $e^-_{ij}$. It will be more intuitive, however, to define two new sets of edges: $\arle{E} \ni \arle{e}_{ij}$ and $\arri{E}\ni \arri{e}_{ij}$ with $\arle{e}_{ij}=e^+_{ij}$ if $v_i<v_j$ and $\arle{e}_{ij}=e^-_{ij}$ if $v_i>v_j$. Likewise, $\arri{e}_{ij}=e^-_{ij}$ if $v_i<v_j$ and $\arri{e}_{ij}=e^+_{ij}$ if $v_i>v_j$. 
We endow the indices with the following structure: $\arri{e}_{ij}\equiv\arle{e}_{ji}$. 

A triplet $(\arle{E},\arri{E},E^0)$ is in one-to-one correspondence with the outer layers of a tubing on a marked graph.
Let us describe this bijection graphically.
One draws a partially directed graph by letting arrows go from $i$ to $j$ for all $\arle{e}_{ij}$, from $k$ to $l$ for all $\arri{e}_{kl}$, and not assigning an orientation to the edge for all $e_{mn}^0$.
Tube all subgraphs made out of $e^0$ edges while intersecting $\arri{e}$ and $\arle{e}$ edges such that the cross is always in the tube that the arrow points to as illustrated in \eqref{eq:GpmTubing}.
Every edge in the graph can be $\rightarrow, \leftarrow$, or $-$, and thus we have $3^E$ functions in total. 
\begin{align}
    \raisebox{-0.35\height}{\includegraphics[height=4.8em]{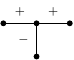}} 
    \hspace{0.1cm}
    \xrightarrow{\hspace{0.8cm}\ }
    \hspace{0.1cm}
    \raisebox{-0.35\height}{\includegraphics[height=4.8em]{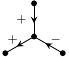}} 
    \hspace{0.1cm}
    \xrightarrow{\hspace{0.8cm}\ }
    \hspace{0.1cm}
    \raisebox{-0.35\height}{\includegraphics[height=4.8em]{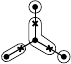}}
    \label{eq:GpmTubing}
\end{align}
Importantly, when $E^0=\emptyset$ this describes how every master integral in \eqref{eq:StudiedIntegrals}, specified by a $\pm$ on every edge, is in one-to-one correspondence with the set of finest CC tube partitions, i.e. tubings with all tubes containing only a single vertex. By dressing this set of $2^E$ tubings with tubed crosses, a bijective relation between time integrals and unparticle tubings is established. The full correlator is thus the sum of all $2^E$ finest unparticle tubings. 

\paragraph{Mergers, swaps, and cuts.}
The admissible tubings on a graph are related by three tube operations: merger, swap, and cut.
In fact, given a single tubing the full set of tubings can be obtained by a sequence of merger and swap operations.
For conformally coupled scalars, these operations act on a tubing in a straightforward way.
Given a CC tubing, with two adjacent tubes $T_1$ and $T_2$, the merger of the tubing with respect to $T_1$ and $T_2$ is a new tubing with $T_1$ and $T_2$ deleted and replaced by the merged tube $T_1 \cup T_2 = (H_1\cup H_2, \partial H_1 \cup \partial H_2 \setminus \partial H_1 \cap \partial H_2, B_1\cup B_2 \setminus \partial H_1 \cap \partial H_2)$.
By adjacent we mean that $T_1$ and $T_2$ cross the same edge but do not share any vertices: $H_1 \cap H_2=\emptyset$ and $\partial H_1 \cap \partial H_2=e_{ij}$.
Merging has a straightforward graphical representation: 
\begin{align}
    \raisebox{-0.35\height}{\includegraphics[width=0.1\linewidth]{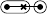}} \xrightarrow{\hspace{0.1cm}\rm merge \hspace{0.1cm}}
    \raisebox{-0.35\height}{\includegraphics[width=0.1\linewidth]{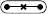}}
\end{align}

The inverse of a merger is obtained by cutting tubes.
Cutting a tube, $T=(H,\partial H,B)$, along an edge $e_{ij}\in H$, is synonymous with partitioning its underlying subgraph $H$ into two smaller CC tubes, $T_1$ and $T_2$.
We can cut along an edge in two ways: either $T_1$ absorbs the cross of $e_{ij}$, so that $B_1\ni e_{ij} \notin B_2$, or the other way around where $B_2 \ni e_{ij} \notin B_1$.
\begin{align}
    \raisebox{-0.43\height}{\includegraphics[width=0.1\linewidth]{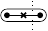}} \xrightarrow{\hspace{0.1cm}\rm cut \hspace{0.1cm}}
    \raisebox{-0.35\height}{\includegraphics[width=0.1\linewidth]{tubings/oper/merg_1.pdf}}
\end{align}

For swap, consider again a CC tubing, and suppose it has two adjacent tubes $T_1$ and $T_2$ where $T_1$ includes the cross $\times_{ij}$ and $T_2$ does not: $e_{ij}\in B_1, e_{ij}\notin B_2$.
The swap of the tubing with respect to $e_{ij}$ is a new tubing with $\times_{ij}$ swapped from $T_1$ to $T_2$: $T_1 \mapsto T_1'$ and $T_2\mapsto T_2'$ with $e_{ij}\notin B_1'$ and $e_{ij}\in B_2'$.
This operation also admits a simple graphical representation.
\begin{align}
    \raisebox{-0.35\height}{\includegraphics[width=0.1\linewidth]{tubings/oper/merg_1.pdf}} \xleftrightarrow{\hspace{0.1cm}\rm swap\hspace{0.1cm}}
    \raisebox{-0.35\height}{\includegraphics[width=0.1\linewidth]{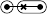}}
\end{align}

These three operations for CC tubings induce their respective counterparts for unparticles in the following way: we first convert the unparticle tubing into a CC tubing by stripping all nested tubes.
We can swap, cut or merge tubes in this tubing using the rules described above.
After operating, one then dresses the CC tubings again adhering to the prescribed arborescence ordering:
\begin{itemize}
    \item \textit{Merger:} strip the unparticle tubing from all its nested tubes, merge two tubes in the resulting CC tubing that are adjacent on an edge $e_{ij}$, and finally dress the resulting CC merger according to the arborescence ordering.
    \begin{align}
        \raisebox{-0.45\height}{\includegraphics[height=4.8em]{tubings/three/three_1.pdf}} \xrightarrow{\hspace{0.1cm}\rm strip \hspace{0.1cm}} \raisebox{-0.45\height}{\includegraphics[height=4.8em]{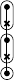}} \xrightarrow{\hspace{0.1cm}\rm merge \hspace{0.1cm}} \raisebox{-0.45\height}{\includegraphics[height=4.8em]{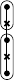}} \xrightarrow{\hspace{0.1cm}\rm dress \hspace{0.1cm}} \raisebox{-0.45\height}{\includegraphics[height=4.8em]{tubings/three/three_7.pdf}}
        \label{eq:GraphMerger}
    \end{align}
    \item \textit{Cut:} strip the unparticle tubing from all its nested tubes, cut the resulting CC tubing along an edge $e_{ij}$, and dress the cut CC tubing again according to the arborescence ordering.
    \begin{align}
        \raisebox{-0.45\height}{\includegraphics[height=4.8em]{tubings/three/three_9.pdf}} \xrightarrow{\hspace{0.1cm}\rm strip \hspace{0.1cm}} \raisebox{-0.45\height}{\includegraphics[height=4.8em]{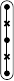}} \xrightarrow{\hspace{0.1cm}\rm cut \hspace{0.1cm}} \raisebox{-0.45\height}{\includegraphics[height=4.8em]{tubings/three/bw/1-3_4-5.pdf}} \xrightarrow{\hspace{0.1cm}\rm dress \hspace{0.1cm}} \raisebox{-0.45\height}{\includegraphics[height=4.8em]{tubings/three/three_7.pdf}}
        \label{eq:GraphCutting}
    \end{align}
    \item \textit{Swap:}
    remove all nested tubes from the unparticle tubing, swap the cross between the tubes $T_1$ and $T_2$ in the resulting CC tubing along the edge $e_{ij}$ on which $T_1$ and $T_2$ are adjacent, nest unparticle tubes in the CC tubing again according to the arborescence ordering.
    \begin{align}
        \raisebox{-0.45\height}{\includegraphics[height=4.8em]{tubings/three/three_7.pdf}} \xrightarrow{\hspace{0.1cm}\rm strip \hspace{0.1cm}} \raisebox{-0.45\height}{\includegraphics[height=4.8em]{tubings/three/bw/1-3_4-5.pdf}} \xrightarrow{\hspace{0.1cm}\rm swap \hspace{0.1cm}} \raisebox{-0.45\height}{\includegraphics[height=4.8em]{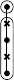}} \xrightarrow{\hspace{0.1cm}\rm dress \hspace{0.1cm}} \raisebox{-0.45\height}{\includegraphics[height=4.8em]{tubings/three/three_8.pdf}}
        \label{eq:GraphSwapping}
    \end{align}
\end{itemize}
Alternatively, we can state the action of the operations directly at the level of the unparticle tubings.
Since unparticle tubings contain sets of nested tubes, the tube operations need to be promoted to maps that act on sets of tubes.
Although it is possible to set up these maps consistently, their action is slightly convoluted and quickly becomes unwieldy.

\subsection{Kinematic flow rules}
\label{sec:rules}
Kinematic flow rules specify the systems of differential equations, $\mathrm{d}\vec{\mathcal I} = A \cdot \vec{\mathcal I}$, by flowing through the space of master integrals.
For each master integral, the right-hand side of its differential equation is described by three quantities: the master integrals or non-zero entries of $A$, the letters that appear on those entries, and the multiplicative factors in terms of vertex twists $\vertextwist_i$ and edge twists $\edgetwist_{ij}$.
Unparticle kinematic flow follows from the sum of four rules:
\begin{equation*}
    \dint \mathcal{I} = \text{activation}\ + \ \text{merger} \ + \ \text{swap} \ + \ \text{copy}.
\end{equation*}
They always produce Abelian flat-connection matrices:
\begin{equation}
    \dint A = 0, \qquad A \wedge A = 0.
\end{equation}

We represent functions in the differential equations by their corresponding graph tubings. 
Square brackets denote $\dint\log$ forms. Given a tube, $T$, and its corresponding letter, $L(T)$, square brackets around a linear function denote the $\dint\log$ form of that function, $[L(T)]\equiv\dint\log L(T)$. When we write square brackets around a tube on a graph they denote the $\dint\log$ form of the corresponding linear function: $[T]\equiv \dint\log L(T)$.
With a slight abuse of terminology, we use ``letter'' to refer to both the linear functions and their $\dint\log$ forms. 
Following the conventions of the previous section, we refer to a tube that is not contained in any other tube as an \textit{outer tube}, and a tube which is enclosed by another tube as a \textit{nested tube}.
The arborescence orders vertices upwards and by extension induces an ordering of the tubes.

The rules are described for the differential of the $i^{\text{th}}$ function, $\dint {\mathcal I}_i$, and we call this function \textit{the differentiated tubing}.
We separate the right-hand side into the functions that appear and the letters, together with their corresponding twist coefficients, that multiply them.
The rules are as follows:
\begin{enumerate}
    \item \textit{Activation.}
    \\
    \begin{itemize}
        \item Function: the differentiated tubing itself. 
        \item 
        Letters: from the tubes contained in the differentiated tubing, add
        \begin{itemize}
        \item every outer tube multiplied by 
        \begin{enumerate}
        \item $\vertextwist_i$ for all vertices $v_i$ inside the tube,
        \item $\edgetwist_{ij}$ for every edge that crosses the tube from below.
        \end{enumerate}
        In the case of the two-site graph, this part of the activation rule gives
        \begin{align*}
        \text{outer tube activation of}\ \ \raisebox{-0.45\height}{\includegraphics[height=1.5cm]{tubings/two/two_1.pdf}} &= \left(\alpha_1\ \left[\raisebox{-0.45\height}{\includegraphics[height=1.35cm]{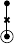}}\right] + (\alpha_2+\edgetwist)\ \left[\raisebox{-0.45\height}{\includegraphics[height=1.35cm]{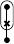}}\right] 
        \right)\ \raisebox{-0.45\height}{\includegraphics[height=1.5cm]{tubings/two/two_1.pdf}} 
        \end{align*}
        If there are multiple edges entering from below, we have to sum the corresponding $\edgetwist_{ij}$:
        \begin{equation*}
            \begin{matrix}
                \text{outer tube} \\
                \text{activation of}
            \end{matrix}
            \raisebox{-0.45\height}{\includegraphics[height=4.8em]{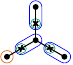}} \ni 
            (\alpha_4+\edgetwist_1+\edgetwist_2)\ \left[\raisebox{-0.45\height}{\includegraphics[height=4.55em]{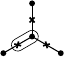}}\right]\raisebox{-0.45\height}{\includegraphics[height=4.55em]{tubings/star/star_4.pdf}}
        \end{equation*}
        \item Add every tubed cross in the tubing with coefficient $\edgetwist_{ij}$. For the two-site graph this entails
        \begin{align*}
        \text{cross activation of}\ \ \raisebox{-0.45\height}{\includegraphics[height=1.5cm]{tubings/two/two_1.pdf}} &= \left(
        \edgetwist\ \left[\raisebox{-0.45\height}{\includegraphics[height=1.35cm]{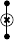}}\right]
        \right)\ \raisebox{-0.45\height}{\includegraphics[height=1.5cm]{tubings/two/two_1.pdf}} 
        \end{align*}
        \item Add every nested tube multiplied by $\edgetwist_{ij}$ corresponding to the up-going edge it crosses. A three-site example illustrates this part of the activation rule clearly:
        \begin{equation*}
            \text{nested tube activation of } \ \raisebox{-0.45\height}{\includegraphics[height=4.8em]{tubings/three/three_6.pdf}} 
            = \edgetwist_2\ \left(\left[\raisebox{-0.45\height}{\includegraphics[height=4.55em]{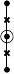}}\right] + \left[\raisebox{-0.45\height}{\includegraphics[height=4.55em]{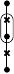}}\right]\right)
        \end{equation*}
    \end{itemize}
    \end{itemize}
    \item \textit{Merger.} \\
    For every pair of adjacent outer tubes $T_\bullet$ and $T_\times$ with $T_\times$ containing the tubed cross $\raisebox{-0.4\height}{\includegraphics[width=4.5em]{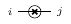}}$:
    \begin{itemize}
        \item Function: strip the differentiated tubing, merge $T_\bullet$ and $T_\times$, and dress the resulting tubing. 
        \item Letters: $T_\times$ minus $T_\bullet$.
        Multiply this difference by a factor of $\edgetwist_{ij}$.
    \end{itemize}
    We have shown how to strip, merge and dress a three-site graph in \eqref{eq:GraphMerger}; let us use the same example to illustrate the merger rule:
    \begin{equation*}
        \text{edge 1 merger of}\ \ \raisebox{-0.45\height}{\includegraphics[height=4.8em]{tubings/three/three_1.pdf}} = \edgetwist_1 \left(\left[\raisebox{-0.45\height}{\includegraphics[height=4.55em]{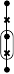}}\right] - \left[\raisebox{-0.45\height}{\includegraphics[height=4.55em]{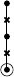}}\right]\right)\ \raisebox{-0.45\height}{\includegraphics[height=4.8em]{tubings/three/three_7.pdf}}
    \end{equation*}
    \item \textit{Swap.} \\
    For every pair of outer tubes $T_1$ and $T_2$ adjacent on the edge $e_{ij}$ and $T_1$ ordered below $T_2$:
    \begin{itemize}
        \item Function: strip the differentiated tubing, swap the cross between $T_1$ and $T_2$, and dress the resulting tubing.
        \item Letters: $\raisebox{-0.4\height}{\includegraphics[width=4.5em]{tubings/cross.pdf}}$ minus $T_2$.
    \end{itemize}
    To exemplify the swap rule, we use the same graph for which we explained the swapping procedure in \eqref{eq:GraphSwapping}:
    \begin{equation*}
        \text{edge 2 swap of} \ \ \raisebox{-0.45\height}{\includegraphics[height=4.8em]{tubings/three/three_7.pdf}}= \edgetwist_2 \left(\left[\raisebox{-0.45\height}{\includegraphics[height=4.55em]{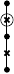}}\right]-\left[\raisebox{-0.45\height}{\includegraphics[height=4.55em]{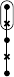}}\right]\right)\ \raisebox{-0.45\height}{\includegraphics[height=4.8em]{tubings/three/three_8.pdf}}
    \end{equation*}
    \item \textit{Copy.} \\
    For every nested tube $T_{\text{nested}}$ inside an outer tube $T_{\text{outer}}$ with an up-going edge $e_{ij}$ enclosed in $T_{\text{outer}}$ and crossing $T_{\text{nested}}$:
    \begin{itemize}
        \item Functions: 
        Strip the differentiated tubing, cut $T_{\text{outer}}$ along $e_{ij}$ where $T_{\text{nested}}$ crosses and dress the resulting tubing; call this function $f_{\text{cut}}$.
        Copy all functions except the differentiated tubing in $\dint f_{\text{cut}}$ on the letter $[T_{\text{nested}}]$.
        \item Letters: $\text{sign}\times([T_{\text{outer}}]-[T_{\text{nested}}])$ where $\text{sign}=-$ if the nested tube $T_{\text{nested}}$ contains the cross on the outgoing edge and $\text{sign}=+$ otherwise.
    \end{itemize}
    As an example, consider the last element of the three-site basis for which we described stripping, cutting and dressing in \eqref{eq:GraphCutting}.
    The copy rule for this particular nested tube reads
    \begin{equation*}
        \text{copy for }T_{\text{nested}}= 
        \raisebox{-0.45\height}{\includegraphics[height=4.55em]{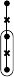}} \ 
        \text{ of } \ 
        \raisebox{-0.45\height}{\includegraphics[height=4.8em]{tubings/three/three_9.pdf}}=\left(\left[\raisebox{-0.45\height}{\includegraphics[height=4.55em]{tubings/three/bw/1-5.pdf}}\right]-\left[\raisebox{-0.45\height}{\includegraphics[height=4.55em]{tubings/three/bw/1-3.pdf}}\right]\right)\ \left((\alpha_1+\alpha_2)\ \raisebox{-0.45\height}{\includegraphics[height=4.8em]{tubings/three/three_7.pdf}} + \alpha_1\ \raisebox{-0.45\height}{\includegraphics[height=4.8em]{tubings/three/three_1.pdf}} - \alpha_1\ \raisebox{-0.45\height}{\includegraphics[height=4.8em]{tubings/three/three_3.pdf}}\right)
    \end{equation*}
where the functions are taken from the differential equation of the cut tubing on $T_{\text{nested}}$:
    \begin{equation*}
        \dint \ 
        \raisebox{-0.45\height}{\includegraphics[height=4.8em]{tubings/three/three_7.pdf}}
        =
        \left((\alpha_1+\alpha_2)\ \raisebox{-0.45\height}{\includegraphics[height=4.8em]{tubings/three/three_7.pdf}} + \alpha_1\ \raisebox{-0.45\height}{\includegraphics[height=4.8em]{tubings/three/three_1.pdf}} - \alpha_1\ \raisebox{-0.45\height}{\includegraphics[height=4.8em]{tubings/three/three_3.pdf}}\right)\left[\raisebox{-0.45\height}{\includegraphics[height=4.55em]{tubings/three/bw/1-3.pdf}}\right]
        \ + \ \text{other letters}
    \end{equation*}
\end{enumerate}

\subsection{Examples}
\label{sec:examples}
In this section, we present the kinematic flow of representative examples.
We present the full result for two-site unparticle exchange, and then discuss the rules applied to selected functions in the three-site chain and four-site star bases.
The full results for the three-site chain and one more example of the four-site star are presented in the appendices~\ref{app:full_KF_3_sites} and~\ref{app:examples_KF_4_star}.

\paragraph{Two-site chain.}
We label the lower vertex by $v_1$ and the upper vertex by $v_2$. The labelling is arbitrary but needs to be consistent for all tubings.
Unlike in kinematic flow for conformally coupled scalars, once the notation is fixed the permutation $v_1 \leftrightarrow v_2$ is no longer a symmetry, since the chosen arborescence ordering breaks it.

The letters in $\dint\log(\cdot)\equiv [\cdot]$ form for the two-site chain are
\begin{align}
    \begin{aligned}
        &\left[\raisebox{-0.45\height}{\includegraphics[height=1.35cm]{tubings/two/bw/1.pdf}}\right] \equiv \left[X_1+Y\right], \quad \left[\raisebox{-0.45\height}{\includegraphics[height=1.35cm]{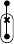}}\right] \equiv \left[X_1-Y\right], \quad \left[\raisebox{-0.45\height}{\includegraphics[height=1.35cm]{tubings/two/bw/2.pdf}}\right] \equiv \left[Y\right], \\
        &\left[\raisebox{-0.45\height}{\includegraphics[height=1.35cm]{tubings/two/bw/2-3.pdf}}\right] \equiv \left[X_2-Y\right], \quad \left[\raisebox{-0.45\height}{\includegraphics[height=1.35cm]{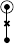}}\right] \equiv \left[X_2+Y\right], \quad \left[\raisebox{-0.45\height}{\includegraphics[height=1.35cm]{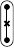}}\right] \equiv \left[X_1+X_2\right].
    \end{aligned}
\end{align}
Applying activation, merger and swap to the first tubing, we obtain
\begin{align}
    \dint\ \raisebox{-0.45\height}{\includegraphics[height=1.5cm]{tubings/two/two_1.pdf}} &= \left(\alpha_1\ \left[\raisebox{-0.45\height}{\includegraphics[height=1.35cm]{tubings/two/bw/1.pdf}}\right] + (\alpha_2+\edgetwist)\ \left[\raisebox{-0.45\height}{\includegraphics[height=1.35cm]{tubings/two/bw/2-3.pdf}}\right] + \edgetwist\ \left[\raisebox{-0.45\height}{\includegraphics[height=1.35cm]{tubings/two/bw/2.pdf}}\right]\right)\ \raisebox{-0.45\height}{\includegraphics[height=1.5cm]{tubings/two/two_1.pdf}} \nonumber \\
    &+\edgetwist\ \left(\left[\raisebox{-0.45\height}{\includegraphics[height=1.35cm]{tubings/two/bw/2-3.pdf}}\right] - \left[\raisebox{-0.45\height}{\includegraphics[height=1.35cm]{tubings/two/bw/1.pdf}}\right]\right)\ \raisebox{-0.45\height}{\includegraphics[height=1.5cm]{tubings/two/two_3.pdf}} + \edgetwist\ \left(\left[\raisebox{-0.45\height}{\includegraphics[height=1.35cm]{tubings/two/bw/2.pdf}}\right] - \left[\raisebox{-0.45\height}{\includegraphics[height=1.35cm]{tubings/two/bw/2-3.pdf}}\right]\right)\ \raisebox{-0.45\height}{\includegraphics[height=1.5cm]{tubings/two/two_2.pdf}}.
    \label{eq:TwoSiteDE1}
\end{align}
The second basis function obeys a similar differential equation: 
\begin{align}
    \dint\ \raisebox{-0.45\height}{\includegraphics[height=1.5cm]{tubings/two/two_2.pdf}} &= \left(\alpha_1\ \left[\raisebox{-0.45\height}{\includegraphics[height=1.35cm]{tubings/two/bw/1-2.pdf}}\right] + (\alpha_2+\edgetwist)\ \left[\raisebox{-0.45\height}{\includegraphics[height=1.35cm]{tubings/two/bw/3.pdf}}\right] + \edgetwist\ \left[\raisebox{-0.45\height}{\includegraphics[height=1.35cm]{tubings/two/bw/2.pdf}}\right]\right)\ \raisebox{-0.45\height}{\includegraphics[height=1.5cm]{tubings/two/two_2.pdf}} \nonumber \\
    &+\edgetwist\ \left(\left[\raisebox{-0.45\height}{\includegraphics[height=1.35cm]{tubings/two/bw/1-2.pdf}}\right] - \left[\raisebox{-0.45\height}{\includegraphics[height=1.35cm]{tubings/two/bw/3.pdf}}\right]\right)\ \raisebox{-0.45\height}{\includegraphics[height=1.5cm]{tubings/two/two_3.pdf}} + \edgetwist\ \left(\left[\raisebox{-0.45\height}{\includegraphics[height=1.35cm]{tubings/two/bw/2.pdf}}\right] - \left[\raisebox{-0.45\height}{\includegraphics[height=1.35cm]{tubings/two/bw/3.pdf}}\right]\right)\ \raisebox{-0.45\height}{\includegraphics[height=1.5cm]{tubings/two/two_1.pdf}}.
    \label{eq:TwoSiteDE2}
\end{align}
Both expressions above contain the common merger function with two nested tubes. In differentiating this function, we need to apply only the activation and copy rules since there are no unmerged edges available,
\begin{align}
    \dint\ \raisebox{-0.45\height}{\includegraphics[height=1.5cm]{tubings/two/two_3.pdf}} &= \left((\alpha_1+\alpha_2)\ \left[\raisebox{-0.45\height}{\includegraphics[height=1.35cm]{tubings/two/bw/1-3.pdf}}\right] + \edgetwist\ \left(\left[\raisebox{-0.45\height}{\includegraphics[height=1.35cm]{tubings/two/bw/1.pdf}}\right] + \left[\raisebox{-0.45\height}{\includegraphics[height=1.35cm]{tubings/two/bw/1-2.pdf}}\right]\right)\right)\ \raisebox{-0.45\height}{\includegraphics[height=1.5cm]{tubings/two/two_3.pdf}} \nonumber \\
    &+\alpha_1\ \left(\left[\raisebox{-0.45\height}{\includegraphics[height=1.35cm]{tubings/two/bw/1-3.pdf}}\right] - \left[\raisebox{-0.45\height}{\includegraphics[height=1.35cm]{tubings/two/bw/1.pdf}}\right]\right)\ \raisebox{-0.45\height}{\includegraphics[height=1.5cm]{tubings/two/two_1.pdf}} + \alpha_1\ \left(\left[\raisebox{-0.45\height}{\includegraphics[height=1.35cm]{tubings/two/bw/1-2.pdf}}\right] - \left[\raisebox{-0.45\height}{\includegraphics[height=1.35cm]{tubings/two/bw/1-3.pdf}}\right]\right)\ \raisebox{-0.45\height}{\includegraphics[height=1.5cm]{tubings/two/two_2.pdf}}.
    \label{eq:TwoSiteDE3}
\end{align}
We can see how the coefficients of the two nested letters are copied from \eqref{eq:TwoSiteDE1} and \eqref{eq:TwoSiteDE2}. For the master integrals at the bottom, this letter appears in a difference with the outer tube so that every function except the differentiated one is multiplied by a \textit{difference} of two letters. 

\paragraph{Three-site chain.} 
Here we present two typical examples from the three-site chain.
We start by enumerating all letters for the three-site chain:
\begin{align}
    &\left[\raisebox{-0.45\height}{\includegraphics[height=4.55em]{tubings/three/bw/1.pdf}}\right] \equiv \left[X_1+Y_{12}\right], 
    &\quad 
    &\left[\raisebox{-0.45\height}{\includegraphics[height=4.55em]{tubings/three/bw/3.pdf}}\right] \equiv \left[X_2+Y_{12}+Y_{23}\right], 
    &\quad 
    &\left[\raisebox{-0.45\height}{\includegraphics[height=4.55em]{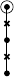}}\right] \equiv \left[X_3+Y_{23}\right], \nonumber \\
    &\left[\raisebox{-0.45\height}{\includegraphics[height=4.55em]{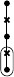}}\right] \equiv \left[X_1-Y_{12}\right], 
    &\quad 
    &\left[\raisebox{-0.45\height}{\includegraphics[height=4.55em]{tubings/three/bw/4-5.pdf}}\right] \equiv \left[X_3-Y_{23}\right], 
    &\quad 
    &\left[\raisebox{-0.45\height}{\includegraphics[height=4.55em]{tubings/three/bw/1-3.pdf}}\right] \equiv \left[X_1+X_2+Y_{23}\right], \nonumber \\
    &\left[\raisebox{-0.45\height}{\includegraphics[height=4.55em]{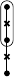}}\right] \equiv \left[X_2+X_3+Y_{12}\right], 
    &\quad 
    &\left[\raisebox{-0.45\height}{\includegraphics[height=4.55em]{tubings/three/bw/2-3.pdf}}\right] \equiv \left[X_2-Y_{12}+Y_{23}\right], 
    &\quad 
    &\left[\raisebox{-0.45\height}{\includegraphics[height=4.55em]{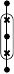}}\right] \equiv \left[X_2-Y_{12}-Y_{23}\right], \nonumber \\
    &\left[\raisebox{-0.45\height}{\includegraphics[height=4.55em]{tubings/three/bw/3-4.pdf}}\right] \equiv \left[X_2+Y_{12}-Y_{23}\right], 
    &\quad 
    &\left[\raisebox{-0.45\height}{\includegraphics[height=4.55em]{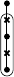}}\right] \equiv \left[X_1+X_2-Y_{23}\right], 
    &\quad 
    &\left[\raisebox{-0.45\height}{\includegraphics[height=4.55em]{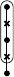}}\right] \equiv \left[X_2+X_3-Y_{12}\right], \nonumber \\
    &\left[\raisebox{-0.45\height}{\includegraphics[height=4.55em]{tubings/three/bw/1-5.pdf}}\right] \equiv \left[X_1+X_2+X_3\right], 
    &\quad 
    &\left[\raisebox{-0.45\height}{\includegraphics[height=4.55em]{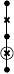}}\right] \equiv \left[Y_{12}\right], 
    &\quad 
    &\left[\raisebox{-0.45\height}{\includegraphics[height=4.55em]{tubings/three/bw/4.pdf}}\right] \equiv \left[Y_{23}\right].
\end{align}
The first example below requires the application of all four rules: the first line applies the activation rule; the first term on the second line is due to a merger while the second term results from the swap rule; and the third and fourth lines are produced by the copy rule.
\begin{align}
    \dint\ \raisebox{-0.45\height}{\includegraphics[height=4.8em]{tubings/three/three_6.pdf}} &= \left(\alpha_1\ \left[\raisebox{-0.45\height}{\includegraphics[height=4.55em]{tubings/three/bw/1-2.pdf}}\right] + (\alpha_2+\alpha_3+\edgetwist_1)\ \left[\raisebox{-0.45\height}{\includegraphics[height=4.55em]{tubings/three/bw/3-5.pdf}}\right] + \edgetwist_1\ \left[\raisebox{-0.45\height}{\includegraphics[height=4.55em]{tubings/three/bw/2.pdf}}\right] + \edgetwist_2\ \left(\left[\raisebox{-0.45\height}{\includegraphics[height=4.55em]{tubings/three/bw/3.pdf}}\right] + \left[\raisebox{-0.45\height}{\includegraphics[height=4.55em]{tubings/three/bw/3-4.pdf}}\right]\right)\right)\ \raisebox{-0.45\height}{\includegraphics[height=4.8em]{tubings/three/three_6.pdf}} \nonumber \\
    &+ \edgetwist_1 \left(\left[\raisebox{-0.45\height}{\includegraphics[height=4.55em]{tubings/three/bw/1-2.pdf}}\right] - \left[\raisebox{-0.45\height}{\includegraphics[height=4.55em]{tubings/three/bw/3-5.pdf}}\right]\right)\ \raisebox{-0.45\height}{\includegraphics[height=4.8em]{tubings/three/three_9.pdf}} + \edgetwist_1 \left(\left[\raisebox{-0.45\height}{\includegraphics[height=4.55em]{tubings/three/bw/2.pdf}}\right]-\left[\raisebox{-0.45\height}{\includegraphics[height=4.55em]{tubings/three/bw/3-5.pdf}}\right]\right)\ \raisebox{-0.45\height}{\includegraphics[height=4.8em]{tubings/three/three_5.pdf}} \nonumber \\
    &+ \left(\left[\raisebox{-0.45\height}{\includegraphics[height=4.55em]{tubings/three/bw/3-5.pdf}}\right]-\left[\raisebox{-0.45\height}{\includegraphics[height=4.55em]{tubings/three/bw/3.pdf}}\right]\right)\ \left((\alpha_2+\edgetwist_1)\ \raisebox{-0.45\height}{\includegraphics[height=4.8em]{tubings/three/three_3.pdf}} - \edgetwist_1\ \raisebox{-0.45\height}{\includegraphics[height=4.8em]{tubings/three/three_7.pdf}} - \edgetwist_1\ \raisebox{-0.45\height}{\includegraphics[height=4.8em]{tubings/three/three_1.pdf}}\right) \nonumber \\
    &+ \left(\left[\raisebox{-0.45\height}{\includegraphics[height=4.55em]{tubings/three/bw/3-4.pdf}}\right]-\left[\raisebox{-0.45\height}{\includegraphics[height=4.55em]{tubings/three/bw/3-5.pdf}}\right]\right)\ \left((\alpha_2+\edgetwist_1)\ \raisebox{-0.45\height}{\includegraphics[height=4.8em]{tubings/three/three_4.pdf}} - \edgetwist_1\ \raisebox{-0.45\height}{\includegraphics[height=4.8em]{tubings/three/three_8.pdf}} - \edgetwist_1\ \raisebox{-0.45\height}{\includegraphics[height=4.8em]{tubings/three/three_2.pdf}}\right)
\end{align}
The second example illustrates multiple applications of the copy rule. Since the outer tube now contains two enclosed edges, the rule must be applied to each of them separately, yielding the second line from the cut along $e_{23}$ and the third and fourth lines from the cut along $e_{12}$.
\begin{align}
    \dint\ \raisebox{-0.45\height}{\includegraphics[height=4.8em]{tubings/three/three_9.pdf}} &= \left((\alpha_1+\alpha_2+\alpha_3)\ \left[\raisebox{-0.45\height}{\includegraphics[height=4.55em]{tubings/three/bw/1-5.pdf}}\right] + \edgetwist_1\ \left(\left[\raisebox{-0.45\height}{\includegraphics[height=4.55em]{tubings/three/bw/1.pdf}}\right] + \left[\raisebox{-0.45\height}{\includegraphics[height=4.55em]{tubings/three/bw/1-2.pdf}}\right]\right) + \edgetwist_2\ \left(\left[\raisebox{-0.45\height}{\includegraphics[height=4.55em]{tubings/three/bw/1-3.pdf}}\right] + \left[\raisebox{-0.45\height}{\includegraphics[height=4.55em]{tubings/three/bw/1-4.pdf}}\right]\right)\right)\ \raisebox{-0.45\height}{\includegraphics[height=4.8em]{tubings/three/three_9.pdf}} \nonumber \\
    &+ \alpha_1 \left(\left[\raisebox{-0.45\height}{\includegraphics[height=4.55em]{tubings/three/bw/1-5.pdf}}\right] - \left[\raisebox{-0.45\height}{\includegraphics[height=4.55em]{tubings/three/bw/1.pdf}}\right]\right)\ \raisebox{-0.45\height}{\includegraphics[height=4.8em]{tubings/three/three_5.pdf}} + \alpha_1 \left(\left[\raisebox{-0.45\height}{\includegraphics[height=4.55em]{tubings/three/bw/1-2.pdf}}\right]-\left[\raisebox{-0.45\height}{\includegraphics[height=4.55em]{tubings/three/bw/1-5.pdf}}\right]\right)\ \raisebox{-0.45\height}{\includegraphics[height=4.8em]{tubings/three/three_6.pdf}} \nonumber \\
    &+ \left(\left[\raisebox{-0.45\height}{\includegraphics[height=4.55em]{tubings/three/bw/1-5.pdf}}\right]-\left[\raisebox{-0.45\height}{\includegraphics[height=4.55em]{tubings/three/bw/1-3.pdf}}\right]\right)\ \left((\alpha_1+\alpha_2)\ \raisebox{-0.45\height}{\includegraphics[height=4.8em]{tubings/three/three_7.pdf}} + \alpha_1\ \raisebox{-0.45\height}{\includegraphics[height=4.8em]{tubings/three/three_1.pdf}} - \alpha_1\ \raisebox{-0.45\height}{\includegraphics[height=4.8em]{tubings/three/three_3.pdf}}\right) \nonumber \\
    &+ \left(\left[\raisebox{-0.45\height}{\includegraphics[height=4.55em]{tubings/three/bw/1-4.pdf}}\right]-\left[\raisebox{-0.45\height}{\includegraphics[height=4.55em]{tubings/three/bw/1-5.pdf}}\right]\right)\ \left((\alpha_1+\alpha_2)\ \raisebox{-0.45\height}{\includegraphics[height=4.8em]{tubings/three/three_8.pdf}} + \alpha_1\ \raisebox{-0.45\height}{\includegraphics[height=4.8em]{tubings/three/three_2.pdf}} - \alpha_1\ \raisebox{-0.45\height}{\includegraphics[height=4.8em]{tubings/three/three_4.pdf}}\right)
\end{align}

\paragraph{Comments on solutions.}
The full correlator is a sum over $2\times 2^E$ in-in sectors.
These sectors come in pairs, with each sector related to its counterpart by complex conjugation.
As discussed above, the remaining $2^E$ independent sectors satisfy differential equations that differ only by sign flips.
By focusing on the all-$+$ in-in sector, we find $3^E$ independent master integrals.
The size of this basis directly controls the number of integration constants, and hence the number of boundary conditions that must be imposed.

The number of master integrals in the all-$+$ sector is the same as in the fully connected sector of conformally coupled wavefunction coefficients \cite{Baumann:2025qjx}.
In that case, the boundary conditions for each sector individually were fixed by imposing the absence of folded singularities, namely the zeros of linear functions associated with tubes containing at least one internal energy with a negative sign, $-Y_{ij}$.
This leaves a single free integration constant in each sector, so that the correlator is a weighted sum of $2^E$ terms.
Imposing the correct factorisation behaviour on partial-energy singularities then mixes the sectors and relates these $2^E$ free coefficients to a single overall factor, thereby uniquely determining the full correlator.
For unparticle exchange, by contrast, all sectors have the same basis size, and the connection matrices do not come naturally in triangular form.
These features may obstruct a direct repetition of the same strategy.
A systematic treatment of boundary conditions in the unparticle case lies beyond the scope of the present work.

An important solution arises when the twist parameters approach zero, $\forall \vertextwist_k,\ \edgetwist_{kl}\to 0$.
In this limit, the underlying physical process is approximated by a tree-level diagram in a power-law cosmology with conformally coupled scalar exchange and cubic interactions.
More precisely, if this situation were exact, then for all vertices $v_k$ and edges $e_{kl}$ in the diagram, \eqref{eq:VertexTwist} and \eqref{eq:EdgeTwist} would give $\vertextwist_k\sim \delta-\epsilon$ and $\edgetwist_{kl}=0$.
In fact, the differential equations for conformally coupled exchange derived in \cite{Baumann:2025qjx} are recovered exactly by rescaling each merger function by $1/\edgetwist_{kl}$ for every edge $e_{kl}$ enclosed by a tube, and then taking the limit $\edgetwist_{kl}\to 0$.
For instance, the two-site differential equations of \cite{Baumann:2025qjx} follow by replacing the merger
\begin{equation}
\raisebox{-0.45\height}{\includegraphics[height=1.5cm]{tubings/two/two_3.pdf}} \
\rightarrow
\frac{1}{\edgetwist} \times \raisebox{-0.45\height}{\includegraphics[height=1.5cm]{tubings/two/two_3.pdf}}
\end{equation}
in \eqref{eq:TwoSiteDE1}, \eqref{eq:TwoSiteDE2}, and \eqref{eq:TwoSiteDE3}, and subsequently setting $\edgetwist=0$, which effectively removes all nested tubes and tubed crosses from the three basis tubings.

Solving the differential equations with all parameters small and generic can then serve as a base case for generating loop-level solutions in which $\alpha_k$ and $\Delta_{kl}$ are close to integer values.
A practical way to implement this is to relate all parameters to a single small variable by setting $\alpha_k=\lambda_k \epsilon$ and $\edgetwist_{kl}=\kappa_{kl} \epsilon$.
As a result, $\epsilon$ factors out of the connection matrix, $A=\epsilon \tilde A$, so that the system takes a form reminiscent of an $\epsilon$-factorised, or canonical, system.
In \cite{Westerdijk:2025ywh}, a similar strategy of relating all twists was applied to the two-site graph.
Crucially, this made it possible to efficiently extract any solution with $\edgetwist_{12}\approx 1,2,\ldots,n$ through the use of shift relations.
Although technically more involved, a similar method may also be applied to the more complicated exchange diagrams considered in this work.

\section{Conclusions and future work}
\label{sec:future}
Kinematic flow for unparticles and loops extends the current state-of-the-art results in a physically interesting direction while uncovering new features in the differential equations.
We studied correlators of conformally coupled scalars exchanging unparticles, or equivalently, banana loops of conformally coupled scalars on power-law FRW backgrounds and mixed massless–conformally coupled banana loops in de Sitter.
The corresponding integrals were most elegantly studied after decomposing the unparticle propagators into two new functions.
We have set up the map between these time integrals and tubings of marked graphs.
To construct these tubings, $E$ ($=$ number of edges) additional tubes need to be nested inside the conformally coupled tubings according to an arbitrarily chosen arborescence ordering of vertices.
Given any tree-level graph exchanging unparticles, the resulting $3^E$ functions constitute a complete and simple basis for the system of differential equations.

The connection matrices are generated by a set of four kinematic flow rules.
Two of them, the activation and merger rules, closely resemble their conformally coupled counterparts with only slight modifications of the twist coefficients. 
The other two are unique to the unparticle systems and describe additional couplings among the basis functions, thereby spoiling the triangular structure of the conformally coupled matrices.
The swap rule does so by adding functions associated with tubings in which a cross on an edge is swapped between two adjacent tubes.
The copy rule breaks apart tubes that contain nested tubes, thereby generating new tubings whose differential equations feed back into that of the original function. 

An interesting application of the unparticle kinematic flow rules would be to study the symbols associated with graphs. 
Recent studies have pointed out that the letters of the conformally coupled systems are related to subsets of cluster variables of certain cluster algebras \cite{Capuano:2025myy,Mazloumi:2025pmx,Paranjape:2026htn,Capuano:2026pgq}.
By imposing a handful of constraints, it is possible to constrain the set of symbols built out of these variables to uniquely bootstrap the result for a given graph \cite{Paranjape:2026htn,Capuano:2026pgq}. 
Remarkably, the new letters, $Y_{ij}$, which play an important role in our work, naturally arise as cluster variables.
This raises the question of whether it is possible to bootstrap unparticle symbols from the perspective of cluster algebras.
In a similar vein, it is natural to ask whether the analytic structure of unparticle exchanges can be captured by a coaction~\cite{McLeod:2026jpz}, whether the basis and differential equations can be constructed geometrically~\cite{Glew:2025ypb,Capuano:2025ehm}, and how the differential equations are related to D-ideals and Gelfand–Kapranov–Zelevinsky (GKZ) systems~\cite{Fevola:2024nzj,Grimm:2025zhv}.

As we have shown, unparticle propagators describe loop integrals of conformally coupled scalars on FRW backgrounds, and additionally massless scalars if we assume de Sitter.
This opens the possibility of computing arbitrarily complicated diagrams built out of two-site banana loops of these two families.
For example, the necklace diagrams \citep[e.g., discussed in][]{Nowinski:2025cvw} made out of massless and conformally coupled scalars can now be treated as multiple tree-level exchange diagrams of unparticles, whose Pfaffian system has already been studied in this work. 
And further it is definitely interesting and challenging to generate the banana loops with arbitrary masses and study their differential structures.

We have outlined a possible strategy for obtaining closed-form solutions, using the differential equations derived in this work to compute a base solution and then applying shift operators to move in the space of twist parameters.
In this way, one may relate the base solution to correlators with different scaling dimensions, including higher-loop examples.
Implementing this strategy requires imposing the absence of folded singularities and the correct factorisation behaviour on partial-energy singularities.
We leave a systematic treatment of these boundary conditions, together with the associated shift relations, to future work.

Although our discussion was restricted to scalar particles, propagators of massless spinning particles do not differ substantially from those of scalars.
As in Feynman integrals, the additional spin structure will introduce a non-trivial ``numerator'' in the loop integrals.
An interesting future avenue of research might be to see whether a type of Passarino--Veltman reduction can be applied to these numerators.

This work adds the unparticle, or banana-loop, case to the two previously understood examples of kinematic flow, conformally coupled and massive exchange.
Taken together, these results suggest that the main exchange processes studied so far admit a unified boundary-centric description in terms of kinematic flow.
It will be interesting to determine how broadly this structure persists in more general settings, and whether it reflects a deeper organising principle.

\section*{Acknowledgements}
Thanks to Daniel Baumann, Harry Goodhew, Austin Joyce, Guilherme Leite Pimentel, and Kamran Salehi Vaziri for comments and insightful discussions.
We thank Guilherme Leite Pimentel for his comments on the draft.
TW thanks Daniel Baumann for the hospitality at UvA and LeCosPA, where part of this work was carried out.

TW and CY are supported by Scuola Normale and by INFN (IS GSS-Pi). 
The research of TW and CY is also supported by the ERC (NOTIMEFORCOSMO, 101126304). 
Views and opinions expressed are, however, those of the author(s) only and do not necessarily reflect those of the European Union or the European Research Council Executive Agency. 
Neither the European Union nor the granting authority can be held responsible for them.

\newpage
\appendix
\section{Derivation of rules}
\label{app:proof_rules}
{In this appendix, we describe a derivation of the kinematic flow rules using the time-integral representation of the basis functions. }

At the level of the time integrals, the first $2^{\text{number of edges}}$ functions in the basis correspond to all possible assignments of $G_\pm$ to the edges of the graph:
\begin{align}
    \mathcal{I}(V,E^+,E^-)=\prod_{v_k \in V}\int \dint \eta_k (-\eta_k)^{-1-\alpha_k} \, e^{i X_k\eta_k}  \prod_{e_{mn}\in E^+} G_{mn}^+ \prod_{e_{mn}\in E^-}G^-_{mn}.
\end{align}
These two propagators, defined in \eqref{eq:GpmDef}, obey a set of partial differential equations. 

The simplest one follows from time-translation symmetry, which implies that the propagators depend only on $\eta_i-\eta_j$, 
\begin{align}
    (\partial_{\eta_i}+\partial_{\eta_j})G^{\pm}_{ij}=0,
    \label{eq:GPDE1}
\end{align}
where we have adopted the shorthand $G^\pm_{ij}\equiv G^\pm_{\edgetwist_{ij}}(Y_{ij};\eta_i,\eta_j)$.

Bessel functions satisfy shift relations which directly imply
\begin{align}
    \partial_{\eta_i}(G^+_{ij}+G^-_{ij})=-iY_{ij}(G^+_{ij}-G^-_{ij}).
    \label{eq:GPDE2}
\end{align}
Finally, the propagators obey two more equations,
\begin{align}
    (\eta_i\partial_{\eta_i}+\eta_j \partial_{\eta_j}-Y_{ij}\partial_{Y_{ij}})G^\pm_{ij}+2\edgetwist_{ij}G_{ij}^\pm &= 0, \\
    (\eta_i-\eta_j)\partial_{\eta_i}G^\pm_{ij}\pm i Y_{ij}(\eta_i-\eta_j)G^\pm_{ij}\pm\edgetwist_{ij}(G_{ij}^\pm-G_{ij}^\mp)&=0,
    \label{eq:GPDE3}
\end{align}
where we have assumed $\varepsilon_{ij}<0$ and non-integer. By analytic continuation, these differential equations also hold for non-integer $\varepsilon_{ij}>0$. 
For the special case of unparticles, time derivatives do not produce any terms proportional to $\delta(\eta_1-\eta_2)$.

This set of partial differential equations can be written as a system of first-order ordinary differential equations by introducing two new functions, $G^0_{ij}\equiv G^0_{\edgetwist_{ij}}(Y_{ij},\eta_i,\eta_j)$ and $\tilde{G}^0_{ij}\equiv \tilde{G}^0_{\edgetwist_{ij}}(Y_{ij},\eta_i,\eta_j)$, which act as sources:
\begin{align}
    (\eta_i\partial_{\eta_i}+ iY_{ij}\eta_i)G^+_{ij}+\lambda \edgetwist_{ij}(G^+_{ij}-G^-_{ij})&=-\edgetwist_{ij}G^0_{ij}\\
    (\eta_j\partial_{\eta_j}-iY_{ij}\eta_j)G^+_{ij}+(1-\lambda) \edgetwist_{ij}(G^+_{ij}-G^-_{ij})&=\edgetwist_{ij}G^0_{ij},\\
    (\eta_i\partial_{\eta_i}- iY_{ij}\eta_i)G^-_{ij}+\tilde \lambda \edgetwist_{ij}(G^-_{ij}-G^+_{ij})&=\edgetwist_{ij}\tilde{G}^0_{ij}\\
    (\eta_j\partial_{\eta_j}+ iY_{ij}\eta_j)G^-_{ij}+(1-\tilde\lambda) \edgetwist_{ij}(G^-_{ij}-G^+_{ij})&=-\edgetwist_{ij}\tilde{G}^0_{ij},
\end{align}
where adding the two equations for $G^+_{ij}$, and likewise those for $G^-_{ij}$, results in the second line of \eqref{eq:GPDE3}.
The two parameters $\lambda$ and $\tilde\lambda$ define a family of different basis choices as they ultimately determine the source functions. 

The choice of $\lambda$ and $\tilde\lambda$ lies at the heart of the unparticle kinematic flow basis and gives rise to the arborescence ordering.
In order to find the simplest structure, we first observe that the shift relation \eqref{eq:GPDE2} relates the equations for $G^+$ with those of $G^-$:
\begin{align}
    (\lambda-\tilde \lambda)(G^+_{ij}-G^-_{ij})=(G^0_{ij}-\tilde{G}^0_{ij}).
\end{align}
Picking $\lambda=\tilde\lambda$ reduces the number of master integrals from $4$ to $3$ since it forces $G^0_{ij}=\tilde{G}^0_{ij}$. 

There are three natural choices for the remaining free parameter $\lambda$.
Either $\lambda=0$ or $\lambda=1$ simplifies the differential equations by removing $G^+_{ij}-G^-_{ij}$ from the $\eta_i\partial_{\eta_i}$ or $\eta_j\partial_{\eta_j}$ equations respectively.
The other possibility, $\lambda=\tfrac12$, leads to the most symmetric form.
We found that an asymmetric choice is preferred for two reasons: the systems of differential equations contain significantly fewer terms, and since the two choices $\lambda=0$ and $\lambda=1$ are related in a straightforward way, it is easy to reconstruct any other choice from those two.

With $\lambda=0$, the system of equations, together with time-translation symmetry, implies two equations for $G^0_{ij}$.
Collecting everything, we have
\begin{align}
    (\eta_i\partial_{\eta_i}+iY_{ij}\eta_i)G^+_{ij}&=-\edgetwist_{ij}G_{ij}^0,\label{eq:FullSystem1}\\
    (\eta_j\partial_{\eta_j}-iY_{ij}\eta_j)G^+_{ij}+\edgetwist_{ij}(G^+_{ij}-G^-_{ij})&=\edgetwist_{ij}G_{ij}^0, \label{eq:FullSystem2}\\
    (\eta_i\partial_{\eta_i}-iY_{ij}\eta_i)G^-_{ij}&=\edgetwist_{ij}G^0_{ij}, \label{eq:FullSystem3}\\
    (\eta_j\partial_{\eta_j}+iY_{ij}\eta_j)G^-_{ij}+\edgetwist_{ij}(G_{ij}^--G^+_{ij})&=-\edgetwist_{ij}G^0_{ij},\label{eq:FullSystem4} \\
    (\eta_i-\eta_j)G^0_{ij}&=\eta_i(G^+_{ij}-G^-_{ij}), \label{eq:FullSystem5}\\
    (\eta_i\partial_{\eta_i}+\eta_j\partial_{\eta_j}+2\edgetwist_{ij})G^0_{ij}&=i Y_{ij}\eta_i(G^+_{ij}+G^-_{ij}) 
    \label{eq:FullSystem6}
    \\
    Y_{ij}(i(\eta_i-\eta_j) + \partial_{Y_{ij}})G_{ij}^+ &= \edgetwist_{ij}(G_{ij}^+ + G_{ij}^-) 
    \label{eq:FullSystem7}
    \\
    Y_{ij}(-i(\eta_i-\eta_j) + \partial_{Y_{ij}})G_{ij}^- &= \edgetwist_{ij}(G_{ij}^+ + G_{ij}^-).
    \label{eq:FullSystem8}
\end{align}
Notice that we can obtain conformally coupled exchange with $\edgetwist_{ij}=0$ by replacing $G^{0}_{ij}\rightarrow \edgetwist_{ij}^{-1} G^{0}_{ij}$ and taking $\edgetwist_{ij}\rightarrow 0$ after multiplying an equation by $\edgetwist_{ij}$ whenever necessary. The resulting system uniquely specifies $G^0_{ij}$ as a collapsing Dirac delta function, thus reproducing the equations found in \cite{Baumann:2025qjx}.

We conclude from the system above that the full basis of master integrals is spanned by
\begin{align}
    \mathcal{I}(V,E^+,E^-,E^0)=\prod_{v_k\in V}\int \dint \eta_k (-\eta_k)^{-1-\alpha_k}e^{iX_k\eta_k} \ G(V,E^+,E^-,E^0),
\end{align}
where
\begin{align}
    G(V,E^+,E^-,E^0)=\prod_{e_{mn}\in E^+}G_{mn}^+\prod_{e_{mn}\in E^-}G_{mn}^-\prod_{e_{mn}\in E^0}G_{mn}^0,
\end{align}
and the sets $E^+$, $E^-$, and $E^0$ partition the full set of edges $E$.
For every edge we have to choose an ordering which specifies the system of equations satisfied by $G_{mn}^\pm$ and $G^0_{mn}$ for that edge.
In other words, for each edge we set either $m=i,n=j$ or $m=j,n=i$ in \eqref{eq:FullSystem1}--\eqref{eq:FullSystem6}.
In the main text, we introduced a consistent way of ordering the vertices using an arborescence and showed how this repartitions the directed edges from $E^+$ and $E^-$ into $\arri{E}$ and $\arle{E}$.
To relate the kinematic flow rules to the time-integral perspective, we adopt exactly the same ordering, but now the ordering is forced upon us if we want to maintain tractable equations.

We have seen how the differential equations for a single propagator simplify when we order $i$ and $j$, i.e. when we order their respective vertices.
We denote $v_m<v_n$ for $m=i,n=j$ and $v_m>v_n$ for $m=j,n=i$.
Products of $G^+$, $G^-$ and $G^0$ therefore require a partial ordering for the set of vertices connected by their associated edges.
In contrast to some generic partial ordering, the arborescence requirement is more subtle but crucial for keeping the differential equations of products of propagators tractable.
We will see this below.
{The specific choice of arborescence ordering does not play a role in the derivation.}

Let $H_i$ be a subgraph of the global graph $G$ containing at least one vertex, and denote the induced edge set $E(H_i)=E^0_i \subseteq E^0$, containing only $e_{ij}^0$ edges. We can write
\begin{align}
\begin{split}
    \left( \left[ \sum_{v_k \in H_i}\eta_k\partial_{\eta_k}+i\eta_k\left( \sum_{\arle{e}_{kl}\in \arle{E}} Y_{kl} - \sum_{\arri{e}_{kl}  \in \arri{E}} Y_{kl}\right) \right] + \sum_{e^0_{mn}\in E_i^0} 2 \edgetwist_{mn} \right) G(V,\arle{E},\arri{E},E^0)= \\
    -i\sum_{e_{mn}^0\in E^0_i} 
    \left\{ \begin{matrix}
        \eta_m & \text{if} \ v_m < v_n \\
        \eta_n & \text{if} \ v_n < v_m 
    \end{matrix}\right\}
    Y_{mn}
    \Big( G(e^0_{mn}\rightarrow \arle{e}_{mn}) + G(e^0_{mn}\rightarrow \arri{e}_{mn}) \Big) \\
    +\sum_{v_k\in V(E^0_i)}\left\{-\sum_{\arle{e}_{kl}\in \arle{E},v_k<v_l}\edgetwist_{kl}G(\arle{e}_{kl}\rightarrow e^0_{kl})+\sum_{\arri{e}_{kl}\in \arri{E},v_l<v_k}\edgetwist_{kl}G(\arri{e}_{kl}\rightarrow e^0_{kl}) \right. \\
    +\sum_{\arle{e}_{kl}\in \arle{E},v_k>v_l}\edgetwist_{kl}\Big[  
    -G-G(\arle{e}_{kl}\rightarrow e^0_{kl})+G(\arle{e}_{kl}\rightarrow \arri{e}_{kl})
    \Big]\\
    \left.
    +\sum_{\arri{e}_{kl}\in \arri{E},v_k>v_l}\edgetwist_{kl}\Big[  
    -G+G(\arri{e}_{kl}\rightarrow e^0_{kl})+G(\arri{e}_{kl}\rightarrow \arle{e}_{kl})
    \Big] \right\},
\end{split}
\end{align}
where the right arrow denotes the replacement of one type of edge with the other. 
Additionally, the product satisfies a simple algebraic equation for $e^0_{ij}$ edges,
\begin{align}
    (\eta_i-\eta_j) G=\eta_i \Big(G(e^0_{ij}\rightarrow \arle{e}_{ij})-G(e_{ij}^0\rightarrow \arri{e}_{ij})\Big)
    \label{eq:AlgTime}
\end{align}
with $v_i<v_j$. 

For the type of time integrals under consideration, integration by parts implies a map between time variable differential operators acting on the integrand and kinematic space operators acting on the integral:
\begin{align}
    \eta_i \cdot G \mapsto -i\frac{\partial}{\partial X_i}\cdot \mathcal{I}, \qquad \eta_i \frac{\partial}{\partial \eta_i}\cdot G \mapsto \left(\vertextwist_i - X_i \frac{\partial}{\partial X_i} \right) \cdot \mathcal{I}.
\end{align}
Hence, we write the integral's differential equation for a generic partition of edges,
\begin{align}
\begin{split}
    \left( \left[ \sum_{v_k \in H_i}\left( X_k - \! \sum_{\arle{e}_{kl}\in \arle{E}} \! Y_{kl}  + \! \sum_{\arri{e}_{kl}  \in \arri{E}} \! Y_{kl}\right) \partial_{X_k} -\alpha_k \right] - \!\!\sum_{e^0_{mn}\in E_i^0}\! 2 \edgetwist_{mn} \right) \mathcal{I}(V,\arle{E},\arri{E},E^0)= \\
    \sum_{e_{mn}^0\in E^0_i} Y_{mn} \cdot
    \left\{ \begin{matrix}
        \partial_{X_m} & \text{if} \ v_m < v_n \\
        \partial_{X_n} & \text{if} \ v_n < v_m 
    \end{matrix}\right\}
    \cdot
    \Big( \mathcal{I}(e^0_{mn}\rightarrow \arle{e}_{mn}) + \mathcal{I}(e^0_{mn}\rightarrow \arri{e}_{mn}) \Big)+\cdots
\end{split}
\end{align}
By invoking the transform of \eqref{eq:AlgTime},
\begin{align}
    \partial_{X_k}\mathcal{I} = \partial_{X_l}\mathcal{I} - \partial_{X_k}\Big( \mathcal{I}(e^0_{kl}\rightarrow \arri{e}_{kl})-\mathcal{I}(e^0_{kl}\rightarrow \arle{e}_{kl}) \Big),
    \label{eq:DifferenceDifferential}
\end{align}
we can iteratively eliminate all $\partial_{X_k}$ derivatives on the left-hand side except a single $\partial_{X_r}$ which corresponds to the vertex that is upper ordered (the local sink in the subarborescence). This is a crucial step which necessitates the use of arborescences: otherwise there would not be a unique vertex that is upper ordered.
\begin{align}
\begin{split}
    \left( \sum_{v_k \in H_i}\left( X_k -\sum_{\arle{e}_{kl}\in \arle{E}} Y_{kl} + \sum_{\arri{e}_{kl}  \in \arri{E}} Y_{kl}\right) \partial_{X_r} -\alpha_k  \right) \mathcal{I}= \\
    \sum_{e_{mn}^0\in E^0_i, \ v_m<v_n} 
    \left(\left[ \sum_{v_j\leq v_m} X_j-\sum_{\arle{e}_{js}\in \arle{E}} Y_{js}+\sum_{\arri{e}_{js}\in\arri{E}}Y_{js} +Y_{mn}\right] \partial_{X_m}
    \mathcal{I}(e^0_{mn}\rightarrow \arri{e}_{mn}) \right. \\ \left.
    -\left[\sum_{v_j\leq v_m} X_j-\sum_{\arle{e}_{js}\in \arle{E}} Y_{js}+\sum_{\arri{e}_{js}\in\arri{E}}Y_{js} -Y_{mn}\right] 
    \partial_{X_m}\mathcal{I}(e^0_{mn}\rightarrow \arle{e}_{mn})+2\edgetwist_{mn}\mathcal{I} \right)
    \\
    +\sum_{v_k\in V(E^0_i)}\left(\sum_{\arle{e}_{kl}\in \arle{E},v_k<v_l}\edgetwist_{kl}\mathcal{I}(\arle{e}_{kl}\rightarrow e^0_{kl})
    -\sum_{\arri{e}_{kl}\in \arri{E},v_k<v_l}\edgetwist_{kl}\mathcal{I}(\arri{e}_{kl}\rightarrow e^0_{kl}) \right. \\
    +\sum_{\arle{e}_{kl}\in \arle{E},v_k>v_l}\edgetwist_{kl}\Big[  
    \mathcal{I}+\mathcal{I}(\arle{e}_{kl}\rightarrow e^0_{kl})-\mathcal{I}(\arle{e}_{kl}\rightarrow \arri{e}_{kl})
    \Big]\\
    \left.
    +\sum_{\arri{e}_{kl}\in \arri{E},v_k>v_l}\edgetwist_{kl}\Big[  
    \mathcal{I}-\mathcal{I}(\arri{e}_{kl}\rightarrow e^0_{kl})-\mathcal{I}(\arri{e}_{kl}\rightarrow \arle{e}_{kl})
    \Big] \right).
    \label{eq:MasterFormula}
\end{split}
\end{align}
Suppose we pick a vertex $v_k\in H_i$ which does not connect to any $e_{kl}^0$ for which $v_l<v_k$. Repeated use of \eqref{eq:DifferenceDifferential} following a chain of ordered vertices lets us write
\begin{align}
    \partial_{X_k} \mathcal{I} = \partial_{X_n} \mathcal{I} - \sum_{v_k \leq v_l < v_m \leq v_n} \partial_{X_l}\Big( \mathcal{I}(e_{lm}^0\rightarrow \arri{e}_{lm})-\mathcal{I}(e_{lm}^0 \rightarrow \arle{e}_{lm}) \Big).
    \label{eq:MasterIteration}
\end{align}

To show how the rules emerge, we only need one more set of differential equations which we obtain by transforming the last two time-variable differential equations, \eqref{eq:FullSystem7} and \eqref{eq:FullSystem8}, to kinematic space,
\begin{align}
    (-\partial_{X_i}+\partial_{X_j}+\partial_{Y_{ij}})\mathcal{I} &= \frac{\edgetwist_{ij}}{Y_{ij}} \Big( \mathcal{I}+\mathcal{I}(\arri{e}_{ij}\rightarrow\arle{e}_{ij}) \Big) & \text{for} \ \arri{e}_{ij} \label{eq:MasterY}
    \\
    (\partial_{X_i}-\partial_{X_j}+\partial_{Y_{ij}})\mathcal{I} &= \frac{\edgetwist_{ij}}{Y_{ij}} \Big( \mathcal{I}+\mathcal{I}(\arle{e}_{ij}\rightarrow\arri{e}_{ij} )\Big) & \text{for} \ \arle{e}_{ij} \\
    \partial_{Y_{ij}}\mathcal{I} &= \partial_{X_i}\Big(\mathcal{I}(e_{ij}^0\rightarrow \arri{e}_{ij})+\mathcal{I}(e_{ij}^0\rightarrow \arle{e}_{ij})\Big) & \text{for} \ e_{ij}^0 .
    \label{eq:MasterY0}
\end{align}

We first set up the map between the master integrals as defined through products of propagators and the tubings of marked graphs.
This is achieved by inspecting what letters appear in the differential equations above.
For any outer layer tube, $T_{\text{outer}}^i$ induced by the subset $H_i$, the linear factor on the left-hand side of \eqref{eq:MasterFormula} multiplying the derivative of the uppermost vertex, $\partial_{X_r}$, corresponds exactly to $L(T^i_{\text{outer}})$.
The vertices and boundary edges of $T_{\text{outer}}^i$ are trivially inherited by those of $H_i$; the $\arle{e}_{kl}$ in the boundary of $H_i$ define $B_{\text{outer}}^i$ of $T_{\text{outer}}^i$ and thus determine the signs of the edge variables $Y_{kl}$. 

To conclude that this is the only letter appearing in \eqref{eq:MasterFormula}, we need to show that every $\partial_{X_m}\mathcal{I}(e^0_{mn}\rightarrow \arle{e}_{mn}/\arri{e}_{mn})$ on the second and third line does not introduce any new letters.
Most importantly, the derivative corresponds to a vertex $v_m$ that is upper ordered within $E_i^0\setminus e^0_{mn}$, i.e. is a local root in the subarborescence.
Examining \eqref{eq:MasterFormula} for these master integrals directly tells us that the letter multiplying $\partial_{X_m}\mathcal{I}(e^0_{mn}\rightarrow \arle{e}_{mn}/\arri{e}_{mn})$ in the $\mathcal{I}$ equation cancels with that arising from the equation for the cut integral $\mathcal{I}(e^0_{mn}\rightarrow \arle{e}_{mn}/\arri{e}_{mn})$.
If the cut tubing does not contain any $e^0_{pm}$ connecting to $v_m$, this suffices.
Otherwise, one has to repeat the argument above until one reaches the last vertices in the ordering, i.e. the leaves of the arborescence. 

The nested tubes can be probed by acting on $\mathcal{I}$ with a differential operator $\partial_{X_i}$ that corresponds to a vertex which is not top-ordered like $v_r$ corresponding to $\partial_{X_r}$ in \eqref{eq:MasterFormula}.
In fact, if we choose a vertex, $v_k$, just below $v_r$, we can conclude from \eqref{eq:MasterIteration} that the letters that $\partial_{X_k}$ picks up consist of the one corresponding to the outer tube and two new ones coming from the outer layers of $\mathcal{I}(e_{kr}^0\rightarrow \arri{e}_{kr}/\arle{e}_{kr})$.
Let us emphasise again that only the outer tube letters of $\mathcal{I}(e_{kr}^0\rightarrow \arri{e}_{kr}/\arle{e}_{kr})$ contribute since $v_k$ is the local root in the subarborescence induced by $E^0_i\setminus e_{kr}^0$ and taking into account the replacement with $\arri{e}_{kr}/\arle{e}_{kr}$.
By going down all paths in the subarborescences of $\mathcal{I}$, one scans through all bipartitions of the outer tubes by cutting along an edge.
The fact that we take derivatives of master integrals with respect to the vertex variables below the cuts always selects the lower tubes to appear in the differential equations.

Two adjacent tubes are connected by either an $\arri{e}_{ij}$ or $\arle{e}_{ij}$ edge.
Without loss of generality, let us take $\arri{e}_{ij}$.
From the above, we know that the tubes and letters for $\partial_{X_i}$ and $\partial_{X_j}$ contain $Y_{ij}$.
This holds even if $v_i$ and/or $v_j$ are inside nested tubes so that there are multiple letters appearing.
The signs of $X_i$ and $Y_{ij}$ in the $\partial_{X_i}$ letters are both positive; for $\partial_{X_j}$, we have $+X_j$ and $-Y_{ij}$.
It is then straightforward to see from \eqref{eq:MasterY} that $\partial_{Y_{ij}}$ introduces no new letters apart from $Y_{ij}$.
On the other hand, \eqref{eq:MasterY0} reveals that derivatives with respect to $Y_{ij}$ for the $e^0_{ij}$ type of edges yield no new letters.

Having shown that the master integrals $\mathcal{I}(V,\arle{E},\arri{E},E^0)$ are in bijection with the tubings on the marked graphs, we merely need to group terms in $\mathcal{I}$'s differential equations to see the rules arise. Let us start by examining the directed edges. 

The \textit{merger rule} emerges when we focus on one $\arri{e}_{kl}$ (or $\arle{e}_{kl}$) at a time.
Since the partition of edges into $(\arle{E},\arri{E},E^0)$ is fixed, the merger $\mathcal{I}(\arri{e}_{kl}\rightarrow e^0_{kl})$ only appears twice in the differential equations no matter how $v_k$ and $v_l$ are ordered: in \eqref{eq:MasterFormula} for a subset $E_k^0$ such that $v_k\in V(E_k^0)$ and for a subset $E_l^0$ with $v_l\in V(E_l^0)$.
Since $\arri{e}_{kl}=\arle{e}_{lk}$, the merger function appears with $-\edgetwist_{kl}$ on the outer-tube letter of $E_k^0$ and with $+\edgetwist_{kl}$ on that of $E_l^0$.
This reproduces exactly the merger rule as stated in the main text.
Notice that the merger function $\mathcal{I}(\arri{e}_{kl}\rightarrow e^0_{kl})$ does not appear anywhere else: only master integrals with zero or more $e^0_{ij}$ edges eliminated from the original $E^0$ feed back into the copy rule. 

The \textit{swap rule} combines \eqref{eq:MasterFormula} and \eqref{eq:MasterY}.
For any directed edge $\arri{e}_{kl}$ or $\arle{e}_{kl}$, the master integral with the edge's orientation flipped, e.g. $\mathcal{I}(\arri{e}_{kl}\rightarrow \arle{e}_{kl})$, will appear with $+\edgetwist_{kl}$ on the $Y_{kl}$ letter.
From the perspective of the marked graphs, reversing the orientation of an edge resembles swapping the cross from one tube to another.
Regardless of the edge's orientation, the $Y_{kl}$ letter is combined with an outer-tube letter determined by the arborescence ordering.
Let $v_k\in V(E_k^0)$ and $v_l \in V(E_l^0)$.
From \eqref{eq:MasterFormula} it is apparent that if $v_k>v_l$, this letter corresponds to the outer layer letter of $E_k^0$, and if $v_k<v_l$, to that of $E_l^0$.
These two subsets of edges are related to outer tubes on the marked graphs and the ordering condition, $v_k \lessgtr v_l$, translates exactly into taking the uppermost tube on the graph. 

The \textit{copy rule} follows from iterating \eqref{eq:MasterIteration} ``maximally'' and substituting \eqref{eq:MasterFormula} for the top-ordered vertex derivative $\partial_{X_n}$.
Generally speaking, there are multiple $E_i^0$ subsets and for each one the induced subarborescence allows for many different paths from the local root (uppermost vertex) down to the leaves (lowest-lying vertices).
The terms in the sum over $e^0_{mn}$ edges on the second and third lines of \eqref{eq:MasterFormula} can be paired up with terms in the sums of all different versions of \eqref{eq:MasterIteration} corresponding to the set of leaves in the arborescence.
Every $e^0_{mn}$ edge corresponds to a pair.
As we already alluded to, the letters in \eqref{eq:MasterFormula} multiplying $\partial_{X_m}\mathcal{I}(e^0_{mn}\rightarrow \arri{e}_{mn}/\arle{e}_{mn})$ exactly cancel the letters in the differential equations of $\mathcal{I}(e^0_{mn}\rightarrow \arri{e}_{mn}/\arle{e}_{mn})$.
Let us call them $\arri{L}_m$ and $\arle{L}_m$ respectively.
The outer-tube letter corresponding to the $E^0_i$ under consideration will be simply denoted by $L_{\text{outer}}$.
A single pair taken from the sums can thus be written as
\begin{align}
    \left( \frac{1}{L_{\text{outer}}}-\frac{1}{\arri{L}_m} \right) \arri{L}_m\partial_{X_m} \mathcal{I}(e_{mn}^0\!\rightarrow \arri{e}_{mn})
    +
    \left( \frac{1}{\arle{L}_m}-\frac{1}{L_{\text{outer}}} \right)  \arle{L}_m\partial_{X_m}\mathcal{I}(e_{mn}^0\!\rightarrow \arle{e}_{mn})+\frac{2\edgetwist_{mn}\mathcal{I}}{L_{\text{outer}}}.
\end{align}
Finally, we have to isolate the original master integral $\mathcal{I}$ in the differential equations of each pair of ``cut functions'' $\partial_{X_m}\mathcal{I}(e^0_{mn}\rightarrow \arri{e}_{mn}/\arle{e}_{mn})$.
From the perspective of the cut functions, $\mathcal{I}$ is a merger and therefore appears with $-\edgetwist_{mn}$ in the expression of $\arri{L}_m\partial_{X_m} \mathcal{I}(e_{mn}^0\rightarrow \arri{e}_{mn})$ and with $+\edgetwist_{mn}$ in that of $\arle{L}_m\partial_{X_m}\mathcal{I}(e_{mn}^0\rightarrow \arle{e}_{mn})$.
The copy rule then corresponds to removing $\mathcal{I}$ from the differential equations of the cut functions,
\begin{align}
    \left( \frac{1}{L_{\text{outer}}}-\frac{1}{\arri{L}_m} \right) \Big[\arri{L}_m\partial_{X_m} &\mathcal{I}(e_{mn}^0\rightarrow \arri{e}_{mn}) + \edgetwist_{mn}\mathcal{I}
    \Big] \nonumber \\
    &+ \left( \frac{1}{\arle{L}_m}-\frac{1}{L_{\text{outer}}} \right) 
    \Big[\arle{L}_m\partial_{X_m}\mathcal{I}(e_{mn}^0\rightarrow \arle{e}_{mn})-
    \edgetwist_{mn}\mathcal{I}
    \Big],
\end{align}
and then setting $\edgetwist_{mn}\mathcal{I}({\arri{L}_m}^{-1}+{\arle{L}_m}^{-1})$ aside for the activation rule.

The \textit{activation rule} simply collects all the contributions proportional to $\mathcal{I}$ itself from all the differential equations.
In this sum, the coefficients of the outer-tube letters can be read off from \eqref{eq:MasterFormula}: $\alpha_k+\edgetwist_{kl}$ for all $l$ such that $v_l<v_k$, i.e. all edges that enter the outer tube from below in the arborescence.
Equations \eqref{eq:MasterY} and \eqref{eq:MasterY0} tell us that $\edgetwist_{ij}/Y_{ij}$ is included for every oriented edge but not for the $e_{ij}^0$ type.
Finally, for every $e_{mn}^0$ in the graph, the feedback of the derivatives of the cut functions contributes $\edgetwist_{mn}\mathcal{I}({\arri{L}_m}^{-1}+{\arle{L}_m}^{-1})$.
With this, we conclude our discussion about how the kinematic flow rules for unparticles and loops follow from differential equations inside the time integrals. 

\section{Kinematic flow of three-site unparticle exchanges}
\label{app:full_KF_3_sites}
In this appendix, we present the full kinematic flow for three-site unparticle exchanges: 
\begin{align}
    \dint\ \raisebox{-0.45\height}{\includegraphics[height=4.8em]{tubings/three/three_1.pdf}} &= \left(\alpha_1\ \left[\raisebox{-0.45\height}{\includegraphics[height=4.55em]{tubings/three/bw/1.pdf}}\right] + (\alpha_2+\edgetwist_1)\ \left[\raisebox{-0.45\height}{\includegraphics[height=4.55em]{tubings/three/bw/2-3.pdf}}\right] + (\alpha_3+\edgetwist_2)\ \left[\raisebox{-0.45\height}{\includegraphics[height=4.55em]{tubings/three/bw/4-5.pdf}}\right] + \edgetwist_1\ \left[\raisebox{-0.45\height}{\includegraphics[height=4.55em]{tubings/three/bw/2.pdf}}\right] + \edgetwist_2\ \left[\raisebox{-0.45\height}{\includegraphics[height=4.55em]{tubings/three/bw/4.pdf}}\right]\right)\ \raisebox{-0.45\height}{\includegraphics[height=4.8em]{tubings/three/three_1.pdf}} \nonumber \\
    &+ \edgetwist_1 \left(\left[\raisebox{-0.45\height}{\includegraphics[height=4.55em]{tubings/three/bw/2-3.pdf}}\right] - \left[\raisebox{-0.45\height}{\includegraphics[height=4.55em]{tubings/three/bw/1.pdf}}\right]\right)\ \raisebox{-0.45\height}{\includegraphics[height=4.8em]{tubings/three/three_7.pdf}} + \edgetwist_1 \left(\left[\raisebox{-0.45\height}{\includegraphics[height=4.55em]{tubings/three/bw/2.pdf}}\right]-\left[\raisebox{-0.45\height}{\includegraphics[height=4.55em]{tubings/three/bw/2-3.pdf}}\right]\right)\ \raisebox{-0.45\height}{\includegraphics[height=4.8em]{tubings/three/three_3.pdf}} \nonumber \\
    &+ \edgetwist_2 \left(\left[\raisebox{-0.45\height}{\includegraphics[height=4.55em]{tubings/three/bw/4-5.pdf}}\right]-\left[\raisebox{-0.45\height}{\includegraphics[height=4.55em]{tubings/three/bw/2-3.pdf}}\right]\right)\ \raisebox{-0.45\height}{\includegraphics[height=4.8em]{tubings/three/three_5.pdf}} + \edgetwist_2 \left(\left[\raisebox{-0.45\height}{\includegraphics[height=4.55em]{tubings/three/bw/4.pdf}}\right] - \left[\raisebox{-0.45\height}{\includegraphics[height=4.55em]{tubings/three/bw/4-5.pdf}}\right]\right)\ \raisebox{-0.45\height}{\includegraphics[height=4.8em]{tubings/three/three_2.pdf}} \\
    \dint\ \raisebox{-0.45\height}{\includegraphics[height=4.8em]{tubings/three/three_2.pdf}} &= \left(\alpha_1\ \left[\raisebox{-0.45\height}{\includegraphics[height=4.55em]{tubings/three/bw/1.pdf}}\right] + (\alpha_2+\edgetwist_1)\ \left[\raisebox{-0.45\height}{\includegraphics[height=4.55em]{tubings/three/bw/2-4.pdf}}\right] + (\alpha_3+\edgetwist_2)\ \left[\raisebox{-0.45\height}{\includegraphics[height=4.55em]{tubings/three/bw/5.pdf}}\right] + \edgetwist_1\ \left[\raisebox{-0.45\height}{\includegraphics[height=4.55em]{tubings/three/bw/2.pdf}}\right] + \edgetwist_2\ \left[\raisebox{-0.45\height}{\includegraphics[height=4.55em]{tubings/three/bw/4.pdf}}\right]\right)\ \raisebox{-0.45\height}{\includegraphics[height=4.8em]{tubings/three/three_2.pdf}} \nonumber \\
    &+ \edgetwist_1 \left(\left[\raisebox{-0.45\height}{\includegraphics[height=4.55em]{tubings/three/bw/2-4.pdf}}\right] - \left[\raisebox{-0.45\height}{\includegraphics[height=4.55em]{tubings/three/bw/1.pdf}}\right]\right)\ \raisebox{-0.45\height}{\includegraphics[height=4.8em]{tubings/three/three_8.pdf}} + \edgetwist_1 \left(\left[\raisebox{-0.45\height}{\includegraphics[height=4.55em]{tubings/three/bw/2.pdf}}\right]-\left[\raisebox{-0.45\height}{\includegraphics[height=4.55em]{tubings/three/bw/2-4.pdf}}\right]\right)\ \raisebox{-0.45\height}{\includegraphics[height=4.8em]{tubings/three/three_4.pdf}} \nonumber \\
    &+ \edgetwist_2 \left(\left[\raisebox{-0.45\height}{\includegraphics[height=4.55em]{tubings/three/bw/2-4.pdf}}\right]-\left[\raisebox{-0.45\height}{\includegraphics[height=4.55em]{tubings/three/bw/5.pdf}}\right]\right)\ \raisebox{-0.45\height}{\includegraphics[height=4.8em]{tubings/three/three_5.pdf}} + \edgetwist_2 \left(\left[\raisebox{-0.45\height}{\includegraphics[height=4.55em]{tubings/three/bw/4.pdf}}\right] - \left[\raisebox{-0.45\height}{\includegraphics[height=4.55em]{tubings/three/bw/5.pdf}}\right]\right)\ \raisebox{-0.45\height}{\includegraphics[height=4.8em]{tubings/three/three_1.pdf}} \\
    \dint\ \raisebox{-0.45\height}{\includegraphics[height=4.8em]{tubings/three/three_3.pdf}} &= \left(\alpha_1\ \left[\raisebox{-0.45\height}{\includegraphics[height=4.55em]{tubings/three/bw/1-2.pdf}}\right] + (\alpha_2+\edgetwist_1)\ \left[\raisebox{-0.45\height}{\includegraphics[height=4.55em]{tubings/three/bw/3.pdf}}\right] + (\alpha_3+\edgetwist_2)\ \left[\raisebox{-0.45\height}{\includegraphics[height=4.55em]{tubings/three/bw/4-5.pdf}}\right] + \edgetwist_1\ \left[\raisebox{-0.45\height}{\includegraphics[height=4.55em]{tubings/three/bw/2.pdf}}\right] + \edgetwist_2\ \left[\raisebox{-0.45\height}{\includegraphics[height=4.55em]{tubings/three/bw/4.pdf}}\right]\right)\ \raisebox{-0.45\height}{\includegraphics[height=4.8em]{tubings/three/three_3.pdf}} \nonumber \\
    &+ \edgetwist_1 \left(\left[\raisebox{-0.45\height}{\includegraphics[height=4.55em]{tubings/three/bw/1-2.pdf}}\right] - \left[\raisebox{-0.45\height}{\includegraphics[height=4.55em]{tubings/three/bw/3.pdf}}\right]\right)\ \raisebox{-0.45\height}{\includegraphics[height=4.8em]{tubings/three/three_7.pdf}} + \edgetwist_1 \left(\left[\raisebox{-0.45\height}{\includegraphics[height=4.55em]{tubings/three/bw/2.pdf}}\right]-\left[\raisebox{-0.45\height}{\includegraphics[height=4.55em]{tubings/three/bw/3.pdf}}\right]\right)\ \raisebox{-0.45\height}{\includegraphics[height=4.8em]{tubings/three/three_1.pdf}} \nonumber \\
    &+ \edgetwist_2 \left(\left[\raisebox{-0.45\height}{\includegraphics[height=4.55em]{tubings/three/bw/4-5.pdf}}\right]-\left[\raisebox{-0.45\height}{\includegraphics[height=4.55em]{tubings/three/bw/3.pdf}}\right]\right)\ \raisebox{-0.45\height}{\includegraphics[height=4.8em]{tubings/three/three_6.pdf}} + \edgetwist_2 \left(\left[\raisebox{-0.45\height}{\includegraphics[height=4.55em]{tubings/three/bw/4.pdf}}\right] - \left[\raisebox{-0.45\height}{\includegraphics[height=4.55em]{tubings/three/bw/4-5.pdf}}\right]\right)\ \raisebox{-0.45\height}{\includegraphics[height=4.8em]{tubings/three/three_4.pdf}}
\end{align}
\begin{align}
    \dint\ \raisebox{-0.45\height}{\includegraphics[height=4.8em]{tubings/three/three_4.pdf}} &= \left(\alpha_1\ \left[\raisebox{-0.45\height}{\includegraphics[height=4.55em]{tubings/three/bw/1-2.pdf}}\right] + (\alpha_2+\edgetwist_1)\ \left[\raisebox{-0.45\height}{\includegraphics[height=4.55em]{tubings/three/bw/3-4.pdf}}\right] + (\alpha_3+\edgetwist_2)\ \left[\raisebox{-0.45\height}{\includegraphics[height=4.55em]{tubings/three/bw/5.pdf}}\right] + \edgetwist_1\ \left[\raisebox{-0.45\height}{\includegraphics[height=4.55em]{tubings/three/bw/2.pdf}}\right] + \edgetwist_2\ \left[\raisebox{-0.45\height}{\includegraphics[height=4.55em]{tubings/three/bw/4.pdf}}\right]\right)\ \raisebox{-0.45\height}{\includegraphics[height=4.8em]{tubings/three/three_4.pdf}} \nonumber \\
    &+ \edgetwist_1 \left(\left[\raisebox{-0.45\height}{\includegraphics[height=4.55em]{tubings/three/bw/1-2.pdf}}\right] - \left[\raisebox{-0.45\height}{\includegraphics[height=4.55em]{tubings/three/bw/3-4.pdf}}\right]\right)\ \raisebox{-0.45\height}{\includegraphics[height=4.8em]{tubings/three/three_8.pdf}} + \edgetwist_1 \left(\left[\raisebox{-0.45\height}{\includegraphics[height=4.55em]{tubings/three/bw/2.pdf}}\right]-\left[\raisebox{-0.45\height}{\includegraphics[height=4.55em]{tubings/three/bw/3-4.pdf}}\right]\right)\ \raisebox{-0.45\height}{\includegraphics[height=4.8em]{tubings/three/three_2.pdf}} \nonumber \\
    &+ \edgetwist_2 \left(\left[\raisebox{-0.45\height}{\includegraphics[height=4.55em]{tubings/three/bw/3-4.pdf}}\right]-\left[\raisebox{-0.45\height}{\includegraphics[height=4.55em]{tubings/three/bw/5.pdf}}\right]\right)\ \raisebox{-0.45\height}{\includegraphics[height=4.8em]{tubings/three/three_6.pdf}} + \edgetwist_2 \left(\left[\raisebox{-0.45\height}{\includegraphics[height=4.55em]{tubings/three/bw/4.pdf}}\right] - \left[\raisebox{-0.45\height}{\includegraphics[height=4.55em]{tubings/three/bw/5.pdf}}\right]\right)\ \raisebox{-0.45\height}{\includegraphics[height=4.8em]{tubings/three/three_3.pdf}} \\
    \dint\ \raisebox{-0.45\height}{\includegraphics[height=4.8em]{tubings/three/three_5.pdf}} &= \left(\alpha_1\ \left[\raisebox{-0.45\height}{\includegraphics[height=4.55em]{tubings/three/bw/1.pdf}}\right] + (\alpha_2+\alpha_3+\edgetwist_1)\ \left[\raisebox{-0.45\height}{\includegraphics[height=4.55em]{tubings/three/bw/2-5.pdf}}\right] + \edgetwist_1\ \left[\raisebox{-0.45\height}{\includegraphics[height=4.55em]{tubings/three/bw/2.pdf}}\right] + \edgetwist_2\ \left(\left[\raisebox{-0.45\height}{\includegraphics[height=4.55em]{tubings/three/bw/2-3.pdf}}\right] + \left[\raisebox{-0.45\height}{\includegraphics[height=4.55em]{tubings/three/bw/2-4.pdf}}\right]\right)\right)\ \raisebox{-0.45\height}{\includegraphics[height=4.8em]{tubings/three/three_5.pdf}} \nonumber \\
    &+ \edgetwist_1 \left(\left[\raisebox{-0.45\height}{\includegraphics[height=4.55em]{tubings/three/bw/2-5.pdf}}\right] - \left[\raisebox{-0.45\height}{\includegraphics[height=4.55em]{tubings/three/bw/1.pdf}}\right]\right)\ \raisebox{-0.45\height}{\includegraphics[height=4.8em]{tubings/three/three_9.pdf}} + \edgetwist_1 \left(\left[\raisebox{-0.45\height}{\includegraphics[height=4.55em]{tubings/three/bw/2.pdf}}\right]-\left[\raisebox{-0.45\height}{\includegraphics[height=4.55em]{tubings/three/bw/2-5.pdf}}\right]\right)\ \raisebox{-0.45\height}{\includegraphics[height=4.8em]{tubings/three/three_6.pdf}} \nonumber \\
    &+ \left(\left[\raisebox{-0.45\height}{\includegraphics[height=4.55em]{tubings/three/bw/2-5.pdf}}\right]-\left[\raisebox{-0.45\height}{\includegraphics[height=4.55em]{tubings/three/bw/2-3.pdf}}\right]\right)\ \left((\alpha_2+\edgetwist_1)\ \raisebox{-0.45\height}{\includegraphics[height=4.8em]{tubings/three/three_1.pdf}} + \edgetwist_1\ \raisebox{-0.45\height}{\includegraphics[height=4.8em]{tubings/three/three_7.pdf}} - \edgetwist_1\ \raisebox{-0.45\height}{\includegraphics[height=4.8em]{tubings/three/three_3.pdf}}\right) \nonumber \\
    &+ \left(\left[\raisebox{-0.45\height}{\includegraphics[height=4.55em]{tubings/three/bw/2-4.pdf}}\right]-\left[\raisebox{-0.45\height}{\includegraphics[height=4.55em]{tubings/three/bw/2-5.pdf}}\right]\right)\ \left((\alpha_2+\edgetwist_1)\ \raisebox{-0.45\height}{\includegraphics[height=4.8em]{tubings/three/three_2.pdf}} + \edgetwist_1\ \raisebox{-0.45\height}{\includegraphics[height=4.8em]{tubings/three/three_8.pdf}} - \edgetwist_1\ \raisebox{-0.45\height}{\includegraphics[height=4.8em]{tubings/three/three_4.pdf}}\right) \\
    \dint\ \raisebox{-0.45\height}{\includegraphics[height=4.8em]{tubings/three/three_6.pdf}} &= \left(\alpha_1\ \left[\raisebox{-0.45\height}{\includegraphics[height=4.55em]{tubings/three/bw/1-2.pdf}}\right] + (\alpha_2+\alpha_3+\edgetwist_1)\ \left[\raisebox{-0.45\height}{\includegraphics[height=4.55em]{tubings/three/bw/3-5.pdf}}\right] + \edgetwist_1\ \left[\raisebox{-0.45\height}{\includegraphics[height=4.55em]{tubings/three/bw/2.pdf}}\right] + \edgetwist_2\ \left(\left[\raisebox{-0.45\height}{\includegraphics[height=4.55em]{tubings/three/bw/3.pdf}}\right] + \left[\raisebox{-0.45\height}{\includegraphics[height=4.55em]{tubings/three/bw/3-4.pdf}}\right]\right)\right)\ \raisebox{-0.45\height}{\includegraphics[height=4.8em]{tubings/three/three_6.pdf}} \nonumber \\
    &+ \edgetwist_1 \left(\left[\raisebox{-0.45\height}{\includegraphics[height=4.55em]{tubings/three/bw/1-2.pdf}}\right] - \left[\raisebox{-0.45\height}{\includegraphics[height=4.55em]{tubings/three/bw/3-5.pdf}}\right]\right)\ \raisebox{-0.45\height}{\includegraphics[height=4.8em]{tubings/three/three_9.pdf}} + \edgetwist_1 \left(\left[\raisebox{-0.45\height}{\includegraphics[height=4.55em]{tubings/three/bw/2.pdf}}\right]-\left[\raisebox{-0.45\height}{\includegraphics[height=4.55em]{tubings/three/bw/3-5.pdf}}\right]\right)\ \raisebox{-0.45\height}{\includegraphics[height=4.8em]{tubings/three/three_5.pdf}} \nonumber \\
    &+ \left(\left[\raisebox{-0.45\height}{\includegraphics[height=4.55em]{tubings/three/bw/3-5.pdf}}\right]-\left[\raisebox{-0.45\height}{\includegraphics[height=4.55em]{tubings/three/bw/3.pdf}}\right]\right)\ \left((\alpha_2+\edgetwist_1)\ \raisebox{-0.45\height}{\includegraphics[height=4.8em]{tubings/three/three_3.pdf}} - \edgetwist_1\ \raisebox{-0.45\height}{\includegraphics[height=4.8em]{tubings/three/three_7.pdf}} - \edgetwist_1\ \raisebox{-0.45\height}{\includegraphics[height=4.8em]{tubings/three/three_1.pdf}}\right) \nonumber \\
    &+ \left(\left[\raisebox{-0.45\height}{\includegraphics[height=4.55em]{tubings/three/bw/3-4.pdf}}\right]-\left[\raisebox{-0.45\height}{\includegraphics[height=4.55em]{tubings/three/bw/3-5.pdf}}\right]\right)\ \left((\alpha_2+\edgetwist_1)\ \raisebox{-0.45\height}{\includegraphics[height=4.8em]{tubings/three/three_4.pdf}} - \edgetwist_1\ \raisebox{-0.45\height}{\includegraphics[height=4.8em]{tubings/three/three_8.pdf}} - \edgetwist_1\ \raisebox{-0.45\height}{\includegraphics[height=4.8em]{tubings/three/three_2.pdf}}\right)
\end{align} 
\begin{align}
    \dint\ \raisebox{-0.45\height}{\includegraphics[height=4.8em]{tubings/three/three_7.pdf}} &= \left((\alpha_1+\alpha_2)\ \left[\raisebox{-0.45\height}{\includegraphics[height=4.55em]{tubings/three/bw/1-3.pdf}}\right] + \alpha_2\ \left[\raisebox{-0.45\height}{\includegraphics[height=4.55em]{tubings/three/bw/4-5.pdf}}\right] + \edgetwist_1\ \left(\left[\raisebox{-0.45\height}{\includegraphics[height=4.55em]{tubings/three/bw/1.pdf}}\right] + \left[\raisebox{-0.45\height}{\includegraphics[height=4.55em]{tubings/three/bw/1-2.pdf}}\right]\right) + \edgetwist_2\ \left[\raisebox{-0.45\height}{\includegraphics[height=4.55em]{tubings/three/bw/4.pdf}}\right]\right)\ \raisebox{-0.45\height}{\includegraphics[height=4.8em]{tubings/three/three_7.pdf}} \nonumber \\
    &+ \edgetwist_2 \left(\left[\raisebox{-0.45\height}{\includegraphics[height=4.55em]{tubings/three/bw/4-5.pdf}}\right] - \left[\raisebox{-0.45\height}{\includegraphics[height=4.55em]{tubings/three/bw/1-3.pdf}}\right]\right)\ \raisebox{-0.45\height}{\includegraphics[height=4.8em]{tubings/three/three_9.pdf}} + \edgetwist_2 \left(\left[\raisebox{-0.45\height}{\includegraphics[height=4.55em]{tubings/three/bw/4.pdf}}\right]-\left[\raisebox{-0.45\height}{\includegraphics[height=4.55em]{tubings/three/bw/4-5.pdf}}\right]\right)\ \raisebox{-0.45\height}{\includegraphics[height=4.8em]{tubings/three/three_8.pdf}} \nonumber \\
    &+ \alpha_1 \left(\left[\raisebox{-0.45\height}{\includegraphics[height=4.55em]{tubings/three/bw/1-3.pdf}}\right]-\left[\raisebox{-0.45\height}{\includegraphics[height=4.55em]{tubings/three/bw/1.pdf}}\right]\right)\ \raisebox{-0.45\height}{\includegraphics[height=4.8em]{tubings/three/three_1.pdf}} + \alpha_1 \left(\left[\raisebox{-0.45\height}{\includegraphics[height=4.55em]{tubings/three/bw/1-2.pdf}}\right] - \left[\raisebox{-0.45\height}{\includegraphics[height=4.55em]{tubings/three/bw/1-3.pdf}}\right]\right)\ \raisebox{-0.45\height}{\includegraphics[height=4.8em]{tubings/three/three_3.pdf}} \\
    \dint\ \raisebox{-0.45\height}{\includegraphics[height=4.8em]{tubings/three/three_8.pdf}} &= \left((\alpha_1+\alpha_2)\ \left[\raisebox{-0.45\height}{\includegraphics[height=4.55em]{tubings/three/bw/1-4.pdf}}\right] + \alpha_2\ \left[\raisebox{-0.45\height}{\includegraphics[height=4.55em]{tubings/three/bw/5.pdf}}\right] + \edgetwist_1\ \left(\left[\raisebox{-0.45\height}{\includegraphics[height=4.55em]{tubings/three/bw/1.pdf}}\right] + \left[\raisebox{-0.45\height}{\includegraphics[height=4.55em]{tubings/three/bw/1-2.pdf}}\right]\right) + \edgetwist_2\ \left[\raisebox{-0.45\height}{\includegraphics[height=4.55em]{tubings/three/bw/4.pdf}}\right]\right)\ \raisebox{-0.45\height}{\includegraphics[height=4.8em]{tubings/three/three_8.pdf}} \nonumber \\
    &+ \edgetwist_2 \left(\left[\raisebox{-0.45\height}{\includegraphics[height=4.55em]{tubings/three/bw/1-4.pdf}}\right] - \left[\raisebox{-0.45\height}{\includegraphics[height=4.55em]{tubings/three/bw/5.pdf}}\right]\right)\ \raisebox{-0.45\height}{\includegraphics[height=4.8em]{tubings/three/three_9.pdf}} + \edgetwist_2 \left(\left[\raisebox{-0.45\height}{\includegraphics[height=4.55em]{tubings/three/bw/4.pdf}}\right]-\left[\raisebox{-0.45\height}{\includegraphics[height=4.55em]{tubings/three/bw/5.pdf}}\right]\right)\ \raisebox{-0.45\height}{\includegraphics[height=4.8em]{tubings/three/three_7.pdf}} \nonumber \\
    &+ \alpha_1 \left(\left[\raisebox{-0.45\height}{\includegraphics[height=4.55em]{tubings/three/bw/1-4.pdf}}\right]-\left[\raisebox{-0.45\height}{\includegraphics[height=4.55em]{tubings/three/bw/1.pdf}}\right]\right)\ \raisebox{-0.45\height}{\includegraphics[height=4.8em]{tubings/three/three_2.pdf}} + \alpha_1 \left(\left[\raisebox{-0.45\height}{\includegraphics[height=4.55em]{tubings/three/bw/1-2.pdf}}\right] - \left[\raisebox{-0.45\height}{\includegraphics[height=4.55em]{tubings/three/bw/1-4.pdf}}\right]\right)\ \raisebox{-0.45\height}{\includegraphics[height=4.8em]{tubings/three/three_4.pdf}} \\  
    \dint\ \raisebox{-0.45\height}{\includegraphics[height=4.8em]{tubings/three/three_9.pdf}} &= \left((\alpha_1+\alpha_2+\alpha_3)\ \left[\raisebox{-0.45\height}{\includegraphics[height=4.55em]{tubings/three/bw/1-5.pdf}}\right] + \edgetwist_1\ \left(\left[\raisebox{-0.45\height}{\includegraphics[height=4.55em]{tubings/three/bw/1.pdf}}\right] + \left[\raisebox{-0.45\height}{\includegraphics[height=4.55em]{tubings/three/bw/1-2.pdf}}\right]\right) + \edgetwist_2\ \left(\left[\raisebox{-0.45\height}{\includegraphics[height=4.55em]{tubings/three/bw/1-3.pdf}}\right] + \left[\raisebox{-0.45\height}{\includegraphics[height=4.55em]{tubings/three/bw/1-4.pdf}}\right]\right)\right)\ \raisebox{-0.45\height}{\includegraphics[height=4.8em]{tubings/three/three_9.pdf}} \nonumber \\
    &+ \alpha_1 \left(\left[\raisebox{-0.45\height}{\includegraphics[height=4.55em]{tubings/three/bw/1-5.pdf}}\right] - \left[\raisebox{-0.45\height}{\includegraphics[height=4.55em]{tubings/three/bw/1.pdf}}\right]\right)\ \raisebox{-0.45\height}{\includegraphics[height=4.8em]{tubings/three/three_5.pdf}} + \alpha_1 \left(\left[\raisebox{-0.45\height}{\includegraphics[height=4.55em]{tubings/three/bw/1-2.pdf}}\right]-\left[\raisebox{-0.45\height}{\includegraphics[height=4.55em]{tubings/three/bw/1-5.pdf}}\right]\right)\ \raisebox{-0.45\height}{\includegraphics[height=4.8em]{tubings/three/three_6.pdf}} \nonumber \\
    &+ \left(\left[\raisebox{-0.45\height}{\includegraphics[height=4.55em]{tubings/three/bw/1-5.pdf}}\right]-\left[\raisebox{-0.45\height}{\includegraphics[height=4.55em]{tubings/three/bw/1-3.pdf}}\right]\right)\ \left((\alpha_1+\alpha_2)\ \raisebox{-0.45\height}{\includegraphics[height=4.8em]{tubings/three/three_7.pdf}} + \alpha_1\ \raisebox{-0.45\height}{\includegraphics[height=4.8em]{tubings/three/three_1.pdf}} - \alpha_1\ \raisebox{-0.45\height}{\includegraphics[height=4.8em]{tubings/three/three_3.pdf}}\right) \nonumber \\
    &+ \left(\left[\raisebox{-0.45\height}{\includegraphics[height=4.55em]{tubings/three/bw/1-4.pdf}}\right]-\left[\raisebox{-0.45\height}{\includegraphics[height=4.55em]{tubings/three/bw/1-5.pdf}}\right]\right)\ \left((\alpha_1+\alpha_2)\ \raisebox{-0.45\height}{\includegraphics[height=4.8em]{tubings/three/three_8.pdf}} + \alpha_1\ \raisebox{-0.45\height}{\includegraphics[height=4.8em]{tubings/three/three_2.pdf}} - \alpha_1\ \raisebox{-0.45\height}{\includegraphics[height=4.8em]{tubings/three/three_4.pdf}}\right)
\end{align}

\section{Examples from kinematic flow of four-site star unparticle exchanges}
\label{app:examples_KF_4_star}
In this appendix, we present some typical differential equations in the kinematic flow of four-site star unparticle exchanges:
\begin{align}
    \dint\ \raisebox{-0.45\height}{\includegraphics[height=4.8em]{tubings/star/star_4.pdf}} &= \left(\alpha_1\ \left[\raisebox{-0.45\height}{\includegraphics[height=4.55em]{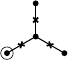}}\right] + \alpha_2\ \left[\raisebox{-0.45\height}{\includegraphics[height=4.55em]{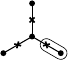}}\right]+ (\alpha_4+\edgetwist_1+\edgetwist_2)\ \left[\raisebox{-0.45\height}{\includegraphics[height=4.55em]{tubings/star/bw/4-5.pdf}}\right]\right. \nonumber \\
    &+ (\alpha_3+\edgetwist_3)\ \left.\left[\raisebox{-0.45\height}{\includegraphics[height=4.55em]{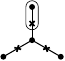}}\right] + \edgetwist_1\ \left[\raisebox{-0.45\height}{\includegraphics[height=4.55em]{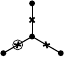}}\right] + \edgetwist_2\ \left[\raisebox{-0.45\height}{\includegraphics[height=4.55em]{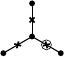}}\right]\right. \nonumber \\
    &+ \edgetwist_3\ \left.\left[\raisebox{-0.45\height}{\includegraphics[height=4.55em]{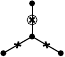}}\right]\right)\ \raisebox{-0.45\height}{\includegraphics[height=4.55em]{tubings/star/star_4.pdf}} \nonumber \\
    &+ \edgetwist_1\ \left(\left[\raisebox{-0.45\height}{\includegraphics[height=4.55em]{tubings/star/bw/4-5.pdf}}\right] - \left[\raisebox{-0.45\height}{\includegraphics[height=4.55em]{tubings/star/bw/1.pdf}}\right]\right)\ \raisebox{-0.45\height}{\includegraphics[height=4.55em]{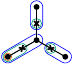}} \nonumber \\
    &+ \edgetwist_1\ \left(\left[\raisebox{-0.45\height}{\includegraphics[height=4.55em]{tubings/star/bw/5.pdf}}\right] - \left[\raisebox{-0.45\height}{\includegraphics[height=4.55em]{tubings/star/bw/4-5.pdf}}\right]\right)\ \raisebox{-0.45\height}{\includegraphics[height=4.55em]{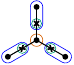}} \nonumber \\
    &+ \edgetwist_2\ \left(\left[\raisebox{-0.45\height}{\includegraphics[height=4.55em]{tubings/star/bw/2-6.pdf}}\right] - \left[\raisebox{-0.45\height}{\includegraphics[height=4.55em]{tubings/star/bw/4-5.pdf}}\right]\right)\ \raisebox{-0.45\height}{\includegraphics[height=4.55em]{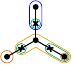}} \nonumber \\
    &+ \edgetwist_2\ \left(\left[\raisebox{-0.45\height}{\includegraphics[height=4.55em]{tubings/star/bw/6.pdf}}\right] - \left[\raisebox{-0.45\height}{\includegraphics[height=4.55em]{tubings/star/bw/4-5.pdf}}\right]\right)\ \raisebox{-0.45\height}{\includegraphics[height=4.55em]{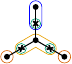}} \nonumber \\
    &+ \edgetwist_3\ \left(\left[\raisebox{-0.45\height}{\includegraphics[height=4.55em]{tubings/star/bw/3-7.pdf}}\right] - \left[\raisebox{-0.45\height}{\includegraphics[height=4.55em]{tubings/star/bw/4-5.pdf}}\right]\right)\ \raisebox{-0.45\height}{\includegraphics[height=4.55em]{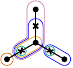}} \nonumber \\
    &+ \edgetwist_3\ \left(\left[\raisebox{-0.45\height}{\includegraphics[height=4.55em]{tubings/star/bw/7.pdf}}\right] - \left[\raisebox{-0.45\height}{\includegraphics[height=4.55em]{tubings/star/bw/3-7.pdf}}\right]\right)\ \raisebox{-0.45\height}{\includegraphics[height=4.55em]{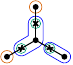}}
\end{align}
\begin{align}
    \dint\ \raisebox{-0.45\height}{\includegraphics[height=4.8em]{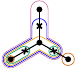}} &= \left((\alpha_1+\alpha_3+\alpha_4+\edgetwist_2)\ \left[\raisebox{-0.45\height}{\includegraphics[height=4.55em]{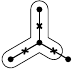}}\right] + \alpha_2\ \left[\raisebox{-0.45\height}{\includegraphics[height=4.55em]{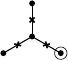}}\right] + \edgetwist_1\ \left[\raisebox{-0.45\height}{\includegraphics[height=4.55em]{tubings/star/bw/1.pdf}}\right]\right. \nonumber \\
    &+ \edgetwist_1\ \left.\left[\raisebox{-0.45\height}{\includegraphics[height=4.55em]{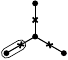}}\right] + \edgetwist_2\ \left[\raisebox{-0.45\height}{\includegraphics[height=4.55em]{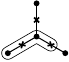}}\right] + \edgetwist_2\ \left[\raisebox{-0.45\height}{\includegraphics[height=4.55em]{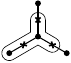}}\right]\right. \nonumber \\
    &+ \edgetwist_2\ \left.\left[\raisebox{-0.45\height}{\includegraphics[height=4.55em]{tubings/star/bw/6.pdf}}\right]\right)\ \raisebox{-0.45\height}{\includegraphics[height=4.55em]{tubings/star/star_22.pdf}} \nonumber \\
    &+ \edgetwist_2\ \left(\left[\raisebox{-0.45\height}{\includegraphics[height=4.55em]{tubings/star/bw/1-5-4-6-7-3.pdf}}\right] - \left[\raisebox{-0.45\height}{\includegraphics[height=4.55em]{tubings/star/bw/2.pdf}}\right]\right)\ \raisebox{-0.45\height}{\includegraphics[height=4.55em]{tubings/star/star_27.pdf}} \nonumber \\
    &+ \edgetwist_2\ \left(\left[\raisebox{-0.45\height}{\includegraphics[height=4.55em]{tubings/star/bw/6.pdf}}\right] - \left[\raisebox{-0.45\height}{\includegraphics[height=4.55em]{tubings/star/bw/1-5-4-6-7-3.pdf}}\right]\right)\ \raisebox{-0.45\height}{\includegraphics[height=4.55em]{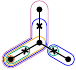}} \nonumber \\
    &+ \alpha_1\ \left(\left[\raisebox{-0.45\height}{\includegraphics[height=4.55em]{tubings/star/bw/1-5-4-6-7-3.pdf}}\right] - \left[\raisebox{-0.45\height}{\includegraphics[height=4.55em]{tubings/star/bw/1.pdf}}\right]\right)\ \raisebox{-0.45\height}{\includegraphics[height=4.55em]{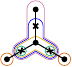}} \nonumber \\
    &+ \alpha_1\ \left(\left[\raisebox{-0.45\height}{\includegraphics[height=4.55em]{tubings/star/bw/1-5.pdf}}\right] - \left[\raisebox{-0.45\height}{\includegraphics[height=4.55em]{tubings/star/bw/1-5-4-6-7-3.pdf}}\right]\right)\ \raisebox{-0.45\height}{\includegraphics[height=4.55em]{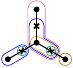}} \nonumber \\
    &+ \left(\left[\raisebox{-0.45\height}{\includegraphics[height=4.55em]{tubings/star/bw/1-5-4-6-7-3.pdf}}\right] - \left[\raisebox{-0.45\height}{\includegraphics[height=4.55em]{tubings/star/bw/1-5-4-6.pdf}}\right]\right)\ \left((\alpha_1+\alpha_4+\edgetwist_2)\ \raisebox{-0.45\height}{\includegraphics[height=4.55em]{tubings/star/star_11.pdf}}\right. \nonumber \\
    &+ \edgetwist_2\ \left.\raisebox{-0.45\height}{\includegraphics[height=4.55em]{tubings/star/star_23.pdf}} - \edgetwist_2\ \raisebox{-0.45\height}{\includegraphics[height=4.55em]{tubings/star/star_9.pdf}} + \alpha_1\ \raisebox{-0.45\height}{\includegraphics[height=4.55em]{tubings/star/star_3.pdf}} - \alpha_1\ \raisebox{-0.45\height}{\includegraphics[height=4.55em]{tubings/star/star_7.pdf}}\right) \nonumber \\
    &+ \left(\left[\raisebox{-0.45\height}{\includegraphics[height=4.55em]{tubings/star/bw/1-5-4-6-7.pdf}}\right] - \left[\raisebox{-0.45\height}{\includegraphics[height=4.55em]{tubings/star/bw/1-5-4-6-7-3.pdf}}\right]\right)\ \left((\alpha_1+\alpha_4+\edgetwist_2)\ \raisebox{-0.45\height}{\includegraphics[height=4.55em]{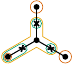}}\right. \nonumber \\
    &+ \edgetwist_2\ \left.\raisebox{-0.45\height}{\includegraphics[height=4.55em]{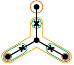}} - \edgetwist_2\ \raisebox{-0.45\height}{\includegraphics[height=4.55em]{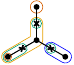}} + \alpha_1\ \raisebox{-0.45\height}{\includegraphics[height=4.55em]{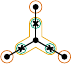}} - \alpha_1\ \raisebox{-0.45\height}{\includegraphics[height=4.55em]{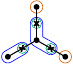}}\right)
\end{align}

\newpage

\addcontentsline{toc}{section}{Reference}
\bibliographystyle{utphys}
{\linespread{1.075}
\bibliography{reference}
}

\end{document}